\documentclass{aa}
\usepackage{longtable}
\usepackage{supertabular}
\usepackage{graphics}
\usepackage{rotating}
\usepackage{natbib}
\begin{document}
%\thesaurus{0.8(0.9.10.1; 0.9.13.2; 0.8.16.5)}
\title{RASS-SDSS Galaxy Clusters Survey.} 
\subtitle{II. A unified picture of the Cluster Luminosity Function.}
\author{P. Popesso\inst{1}, H. B\"ohringer\inst{1}, M. Romaniello\inst{2}, W. Voges\inst{1}}
\institute{ Max-Planck-Institut fur extraterrestrische Physik, 85748 Garching, Germany
\and European Southern Observatorty,  Karl Scharzschildstr. 2, 85748, ESO, ESO}
\authorrunning{P. Popesso et al.}
\maketitle

\begin{abstract}
We constructed the composite luminosity function (LF) of clusters of
galaxies in the five SDSS photometric bands u,g,r,i and z from the
RASS-SDSS galaxy cluster catalog. Background and foreground galaxies
are subtracted using both a local and a global background correction
to take in account the presence of large scale structures and
variations from field to field, respectively. The composite LF clearly
shows two components: a bright-end LF with a classical slope of -1.25
in each photometric band, and a faint-end LF much steeper ($-2.1 \le
\alpha \le  -1.6$) in the dwarf  galaxy region. The observed upturn of
the faint galaxies has a location ranging from -16 +5log(h) in the g
band to -18.5 +5log(h) in the z band.  To study the universality of
the cluster LF we compare the individual cluster LFs with the
composite luminosity function. We notice that, in agreement with the
composite LF, a single Schechter component is not a good fit for the
majority of the clusters.  We fit a Schechter function to the
bright-end of the individual clusters LFs in the magnitude region
brighter than the observed upturn of the dwarf galaxies. We observe
that the distributions of the derived parameters is close to a
Gaussian around the value of the composite bright-end LF parameters
with a dispersion compatible with the statistical errors. We conclude
that the bright-end of the galaxy clusters is universal.  To study the
behavior of the individual faint-end LF we define the Dwarf to Giant
galaxy Ratio (DGR) of the single clusters. We notice that the
distribution of DGR has a spread much larger than the statistical
errors. Our conclusion is that the cluster luminosity function is not
universal since the cluster faint-end, differently from the
bright-end, varies from cluster to cluster.
\end{abstract}

\section{Introduction}
The galaxy luminosity function (LF) is one of the most direct
observational test of theories of galaxy formation and evolution.
Clusters of galaxies are ideal systems within which to measure the
galaxy LF for the large number of galaxies at the same distance. There
are two main purposes in the study of the cluster LF: the comparison
of the galaxy LF in clusters and field and thus the study of the
influence of the environment on the global statistical properties of
galaxies, and the search for differences in the LF of different
clusters as indicators of differences in the galaxy formation due to
environmental effects or dynamical processes.

The cluster galaxy over-density with respect to the surrounding field
is sufficiently high to efficiently identify members either
photometrically through the statistical removal of foreground and
background galaxies or spectroscopically. These techniques have been
used to measure LFs for individual clusters or to form a composite LF,
in order to eliminate the peculiarity of the individual LFs and
enhance the underlying possibly universal LF (Dressler 1978; Lugger
1986; Colless 1989; Lugger 1989; Lumsden et al.  1997; Valotto et
al. 1997; Rauzy et al.  1998; Garilli et al.  1999; Paolillo et
al. 2001; Goto et al.  2002; Yagi et al. 2002).  Many of these studies
do not agree on the exact form of the LF.  Several authors (Dressler
1978; Lumdsen 1997; Valotto et al.  1997; Garilli et al. 1999; Goto et
al. 2002) found differences between the LFs of different clusters and
between cluster and field, while others (Lugger 1986; Colless 1989;
Lugger 1989; Rauzy 1998; Trentham 1998; Paolillo et al.  2001)
concluded that the galaxy LF is universal in all
environments. However, all these works used different techniques
to check the universality of the cluster LF. Therefore, it is
difficult to understand if their conclusions depend on the different
tests beeing applied or to actual physical distinctions. Table 1
summarizes the variations between previous studies in the same color
and their $\sigma$ error limits for the Schechter parameters $M^*$ and
$\alpha$.  We have transformed magnitudes to $H_0=100$ km
$\rm{s}^{-1}$ $\rm{Mpc} ^{-1}$ without changing their cosmology.

So far, the majority of the studies on the cluster composite LF has
concentrated on the slope at the relatively bright end of the cluster
LF ($M_g \le -17$) without taking into account the behavior of the
dwarf galaxy population in clusters. Instead much work has been done
in the recent years in measuring the faint end ($-18 \le M_g \le -10$)
of the galaxy LF in several nearby clusters (e.g.  Driver 1994; Smith
et al.  1997; Phillipps et al.  1998; Boyce et al.  2001;
Beijersbergen et al.  2001; Sabatini et al.  2002; Trentham 2003;
Cortese et al.  2004).  The LF of these clusters typically steepens
faintward of about $M_g \sim -18$ by showing the debated upturn of the
dwarf galaxies .  The faint end slope $\alpha$ of the LF in this range
of magnitudes typically lies in the range -1.4 to -2.2.  Phillipps et
al. 1998 noted that the steepness of the faint end slope appears to
depend on the cluster density, with dwarfs being more common in lower
density environments.  This is possibly because the various dynamical
processes which can destroy dwarf galaxies act preferentially in dense
environments.

In this paper we present the analysis of the cluster composite LF
based on the second release of the Sloan Digital Sky survey (SDSS DR2,
Abazajian et al.  2004). The excellence of the SDSS DR2 in terms of its
size, depth and sky coverage and the accurate photometry in 5
different optical wavebands gives unprecedented advantages in
comparison to the previous studies.  Firstly, the sky coverage (3324
$\rm{deg}^2$) gives us the possibility to overcome the well known
problem of the statistical subtraction of the galaxy background. We
used large areas of the survey to define a mean global galaxy
background and a region close to the clusters to determine the local
galaxy background in order to check for systematics in the field
subtraction. Secondly, the apparent magnitude limit of the SDSS DR2 in
all the five bands is sufficiently deep (e.g.  $r_{lim}=22.2$, 95\%
completeness) that, at the mean redshift of our cluster sample ($z
\sim 0.15$), the cluster LF can extend and cover a significant part of
the dwarf region, going deeper than in all previous studies of the
composite luminosity function (more than 6 magnitudes fainter than
M*). Thirdly, the high accuracy of the SDSS photometry in all bands
gives us the possibility to measure in a statistically significant way
the individual cluster LF with the consequent opportunity to check
directly the universality of the LF. Furthermore, the accurate
multi-color photometry allows us to use several objectively-measured
galaxy properties like galaxy morphology. Finally, our comparison of
the cluster and field LFs can done  within the SDSS data.

To calculate the cluster composite LF we used the RASS-SDSS galaxy
cluster sample (Popesso et al. 2004), which includes 130 systems
observed in X-rays.  The use of the RASS-SDSS galaxy cluster catalog
ensures that none of the systems is a simple projection of large scale
structure along the line of sight.

In this paper we focus on the bright end of the cluster LF while
a more detailed analysis of the faint end will appear in a fortcoming
paper. The paper is organized as follows: in sect.  2 we describe
the properties of the cluster sample and the optical galaxy
photometry; in sect. 3 we explain the methods used in constructing the
individual cluster LFs and the methods of the background subtraction,
in sect. 4 we describe the methods used for building the Composite LF,
in sect. 5 we describe in details our results and finally sect. 6
contains our conclusions.  Throughout the paper we use $\rm{H}_0=100$
$\rm{km}$$\rm{s}^{-1}$$\rm{Mpc}^{-1}$, $\Omega_{m}=0.3$ and
$\Omega_{\lambda}=0.7$.

\begin{table*}
\begin{center}
\begin{tabular}[b]{cccccc}
\hline
Reference & $M^*$ & $\alpha$ & Band & Ncluster& Luminosity range \\
\hline
Goto et al. 2002 & -20.84 $\pm$0.26 & -1.40$\pm$0.11 & u& 204 & $-24 \le M_u \le -18$ \\
\hline
\hline
Schechter (1976) &-19.9$\pm$0.50 & -1.24 & $b_j$ & 13 & $-22.5 \le M_J \le -18.5$ \\
Dressler (1978)& -19.7$\pm$0.50 & -1.25 & F & 12 & $-23.5 \le M_F \le -18.5$ \\
Colless (1989) & -20.10$\pm$0.07 & -1.25 & $b_j$ & 14 & $-22.5 \le M_J \le -17$ \\
Lumsden et al. (1997) & -20.16$\pm$0.02 & -1.22$\pm$0.04 & $b_j$ & 46 & $-21 \le M_b \le -18$ \\
Valotto et al. (1997) & -20.00$\pm$0.10 & -1.40$\pm$0.10& $b_j$ & 55 & $-21 \le M_b \le -17$ \\
Rauzy et al. (1998) & -20.91$\pm$0.21 & -1.50$\pm$0.11 & $b_j$ & 28 & $-21 \le M_b \le -17$ \\
Garilli et al. (1999) & -20.30$\pm$0.10 & -0.94$\pm$0.07 & g & 65 & $-22.5 \le M_g \le -15.5$ \\
Paolillo et al. (2001) & -20.22$\pm$0.15 & -1.07$\pm$0.08 & g & 39 & $-24.5 \le M_g \le -16.5$ \\
Goto et al. (2002) & -21.24$\pm$0.11 &  -1.00$\pm$0.06 & g& 204 & $-24 \le M_g \lg -18$ \\
De Propris et al. (2003) & -20.07$\pm$0.07& -1.28$\pm$0.03 & $b_j$& 60 & $-22.5 \le M_b \le -16$ \\
\hline
\hline
Lugger et al (1989) & -21.31$\pm$0.13 & -1.21$\pm$0.09 & R & 9& $-23 \le M_R \le -18.5$ \\
Garilli et al. (1999) & -20.66$\pm$0.16 & -0.95$\pm$0.07 & r & 65 & $-22.5 \le M_r \le -15.5$ \\
Paolillo et al. (2001) & -20.67$\pm$0.16 & -1.11$\pm$0.08 & r & 39 & $-24.5 \le M_r \le -16.5$ \\
Yagi et al. (2002) & -21.30$\pm$0.20 & -1.31$\pm$0.05 & $R_C$ & 10 & $-23.5 \le M_Rc \le -16$ \\
Goto et al. (2002) & -21.44$\pm$0.05 & -0.85$\pm$0.03 & r & 204 & $-24 \le M_r \le -18$ \\
\hline
\hline
Paolillo et al. (2002) & -20.85$\pm$0.20& -1.09$\pm$0.11& i & 39 & $-24 \le M_i \le -17$ \\
Goto et al. (2002) & -21.54$\pm$0.08 & -0.70$\pm$0.05 & i & 204 & $-24 \le M_i \le -18$ \\
\hline
\hline
Goto et al. (2002) & -21.59$\pm$0.06 & -0.58$\pm$0.04 & z & 204 & $-24 \le M_z \le -18$ \\
\hline
\hline
\end{tabular}
\caption{ Schechter parameters fitted to the Composite LF retrieved in the literature.}
\end{center}
\end{table*}

\section{The data}
The RASS-SDSS galaxy cluster catalog comprises 130 systems detected in
the ROSAT All Sky Survey (RASS).  The X-ray cluster properties and the
cluster redshift have been taken from different X-ray catalogs: the
ROSAT-ESO flux limited X-ray cluster sample (REFLEX, B\"ohringer et
al.  2003), the Northern ROSAT All-sky cluster sample (NORAS,
B\"ohringer et al.  2000), the NORAS 2 cluster sample (Retzlaff 2001),
the ASCA Cluster Catalog (ACC) from Horner et al. (2001) and the Group
Sample (GS) of Mulchaey et al.  2003.  In constructing the composite
LF we restricted our selection to clusters with $z \le 0.25$ in order
to sample well below the predicted M*, and used therefore 97 clusters
of 130 systems in the catalog.

The optical photometric data are taken from the SDSS DR2 (York et
al. 2000, Stoughton et al.  2002 and Abazajian et al.  2004).  The
SDSS consists of an imaging survey of $\pi$ steradians of the northern
sky in the five passbands u, g, r ,i, z, in the entire optical range
from the atmospheric ultraviolet cutoff in the blue to the sensitivity
limit of silicon in the red.  The survey is carried out using a 2.5 m
telescope, an imaging mosaic camera with 30 CCDs, two fiber-fed
spectrographs and a 0.5 m telescope for the photometric calibration.
The imaging survey is taken in drift-scan mode.  The imaging data are
processed with a photometric pipeline (PHOTO) specially written for
the SDSS data.  For each cluster we defined a photometric galaxy
catalog as describe in section 3 of Popesso et al. 2004.

\begin{figure}
\begin{center}
\begin{minipage}{0.5\textwidth}
\resizebox{\hsize}{!}{\includegraphics{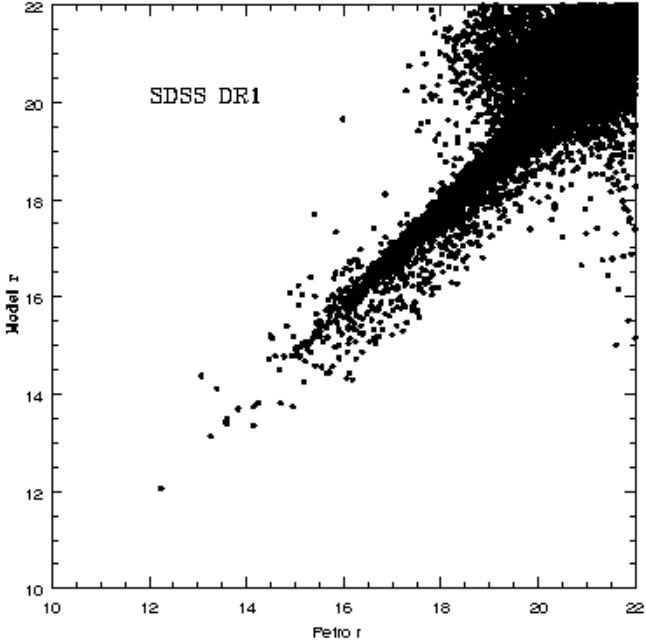}}
\end{minipage}
\end{center}
\caption{
Petrosian magnitude versus Model magnitude in the Data Release 1 (DR1).}
\label{dr1}
\end{figure}

\begin{figure}
\begin{center}
\begin{minipage}{0.5\textwidth}
\resizebox{\hsize}{!}{\includegraphics{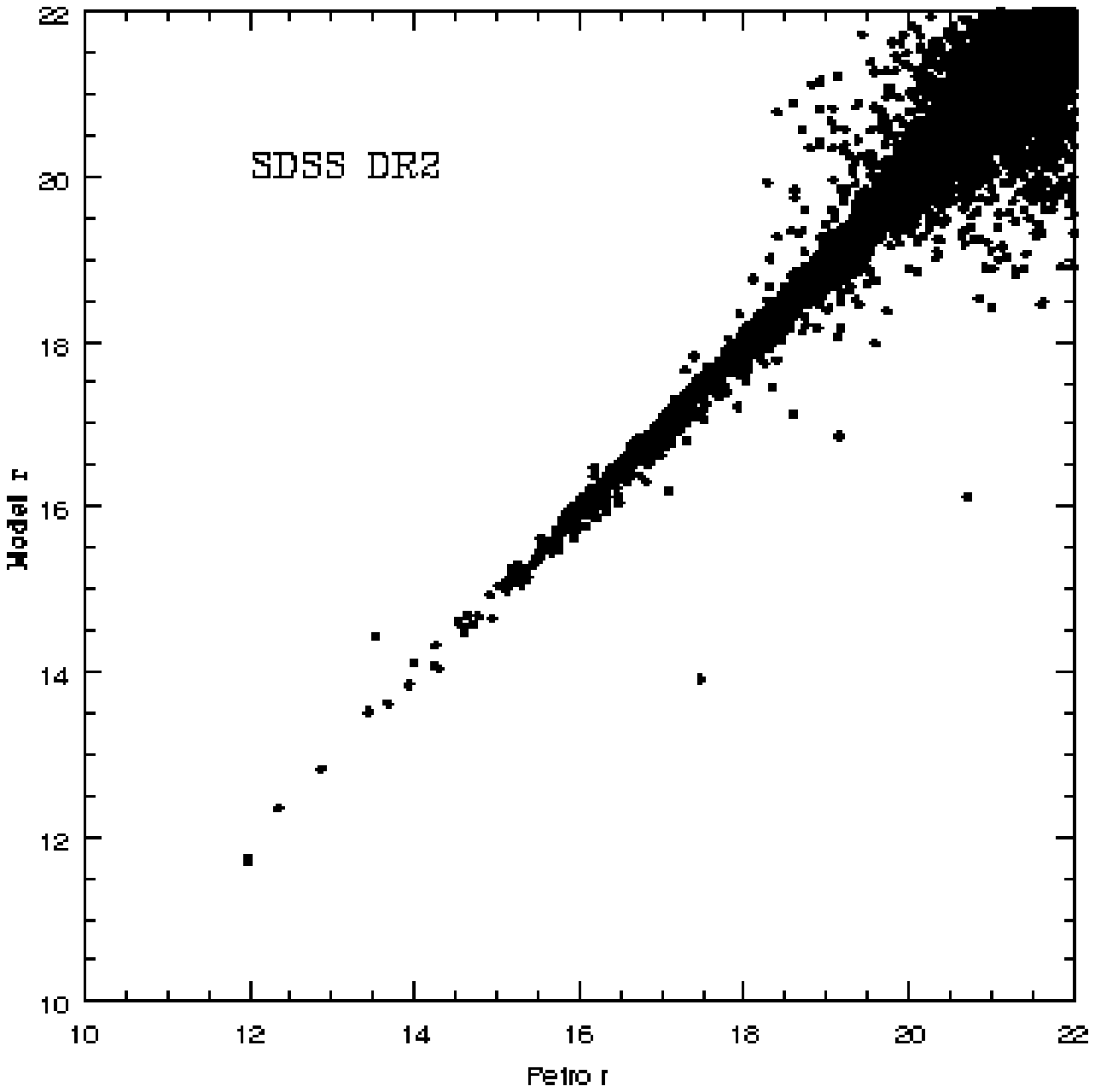}}
\end{minipage}
\end{center}
\caption{
Petrosian magnitude versus Model magnitude in the Data Release 2 (DR2).}
\label{dr2}
\end{figure}

For the analysis in this paper we use only SDSS Model magnitudes. Due
to a bug of PHOTO, found during the completion of DR1, the model
magnitudes are systematically under-estimated by about 0.2-0.3
magnitudes for galaxies brighter then 20th magnitude, and accordingly
the measured radii are systematically too large. This problem has
been fixed in the SDSS DR2, therefore the model magnitude can be
considered a good estimate of the galaxy total luminosity at any
magnitude and are not dependent on the seeing as the Petrosian
magnitudes. Figs.  \ref{dr1} and \ref{dr2} show the difference in the
quality of the galaxy photometry between the DR1 and the DR2 data. For
this study we only use the revised DR2 for the complete cluster
sample.

\section{The individual Luminosity Functions}

\subsection{Background subtraction}

We consider two different approaches to the statistical
subtraction of the galaxy background.  First we calculate a
local background in an annulus with inner radius of 3 Mpc
$\rm{h}^{-1}$ from the X-ray cluster center and width of 0.5 deg.  The
annulus is divided in 20 sectors ( Popesso et al. 2004) and
those featuring a larger than $3\sigma$ deviation from the median
galaxy density are discarded from the further calculation.  In this
way other clusters close to the target or voids are not included in
the background correction.  We compute the galaxies number
counts $N_{bg}^l(m)dm$ per bin of magnitude (with a bin width of 0.5
mag) and per squared degree in the remaining area of the annulus.  The
statistical source of error in this approach is the Poissonian
uncertainty of the counts, given by $\sqrt{(N_{bg}^l(m))}$.

As a second method we derive a global background correction. The
galaxy number counts $N_{bg}^g(m)dm$ is derived from the mean of the
magnitude number counts determined in five different SDSS sky regions,
each with an area of 30 $\rm{deg^2}$.  The source of uncertainty in
this second case is systematic and originates the presence of
large-scale clustering within the galaxy sample, while the Poissonian
error of the galaxy counts is small due to the large area involved. We
estimate this error as the standard deviation of the mean global
number counts, $\sigma_{bg}^g(m)$, in the comparison of the five
areas.  To take into account this systematic source of error also for
the the local background , we estimate the background number
counts error as
$\sigma_{bg}(m)=max(\sqrt{(N_{bg}^l(m))},\sigma_{bg}^g(m))$ (Lumdsen
et al. 1997) for all the derived quantities. For a detailed comparison
of the results of the local and global background estimates see
Popesso et al. (2004).

\subsection{Luminosity Function}
We derive the individual cluster luminosity function by subtracting
from the galaxy counts measured in a certain region the local or the
global field counts rescaled to the cluster area.  We calculate the
individual cluster LF within different radii, from 0.3 to 2 Mpc
$\rm{h}^{-1}$, to study possible dependences of the LF on the
clustercentric distance and thus on the density. According to the work
in the literature, we exclude from the individual cluster LFs the
Brightest Cluster Galaxies (BCG).

To build  the Composite   LF we transform  the apparent   magnitude in
absolute magnitude according to:
\begin{equation}
M=m - 25 -5log_{10}(D_L/1Mpc) - A - K(z)
\end{equation}
where $D_L$ is the luminosity distance, A is the Galactic extinction
and $K(z)$ is the K-correction.  We deredden the Petrosian and model
magnitudes of galaxies using the Galactic map of Schlegel et al.
(1998) in each photometric band.  We use the K-correction supplied by
Fukugita, Shimasaku, $\&$ Ichikawa (1995) for elliptical galaxies,
assuming that the main population of our clusters are the old
elliptical galaxies at the cluster redshift.

Due to the high accuracy of the SDSS multi-color photometry, the
quality of the individual cluster LF is very high. Therefore, to
compare the Schechter parameters of the individual LF with those of
the composite luminosity function, we fit a Schechter luminosity
function to the single clusters by using the fitting method described
in the section 4 of Popesso et al. 2004.

\section{The Composite Luminosity Function}
The composite LF is not only a good method to calculate with high
accuracy the cluster LF when the quality of the individual cluster LFs
is too low, but it is also a tool to check for the LF
universality. The composite LF can be easily interpreted as a mean
cluster LF.  Therefore, the distribution of the individual LF
parameters should be Gaussian around the corresponding value of the
Composite LF parameters, if the LF is universal in all the cluster
environments.  A good description for the calculation of the composite
LF can be found in Colless (1989).  Following these prescriptions, the
Composite LF is built by summing the cluster galaxies in absolute
magnitude bins and scaling by the richness of their parent clusters:
\begin{equation}
N_{cj}=\frac{N_{c0}}{m_j}\sum_i{\frac{N_{ij}}{N_{i0}}}
\end{equation}
where $N_{cj}$ is the number of galaxies in the $j\rm{th}$ absolute
magnitude bin of the composite LF, $N_{ij}$ is the number in the
$j\rm{th}$ bin of the $i\rm{th}$ cluster LF, $N_{i0}$ is the
normalization used for the $i\rm{th}$ cluster LF, $m_j$ is the number
of clusters contributing to the $j\rm{th}$ bin and $N_{c0}$ is the sum
of all the normalizations:
\begin{equation}
N_{c0}=\sum_i{{N_{i0}}}.
\end{equation}
Since all the systems in the cluster sample cover the magnitude region
$M \le -19$ in the five wavebands, we choose that region for the
normalization  according to the treatment in the literature.

\begin{figure*}
\begin{center}
\begin{minipage}{0.8\textwidth}
\resizebox{\hsize}{!}{\includegraphics{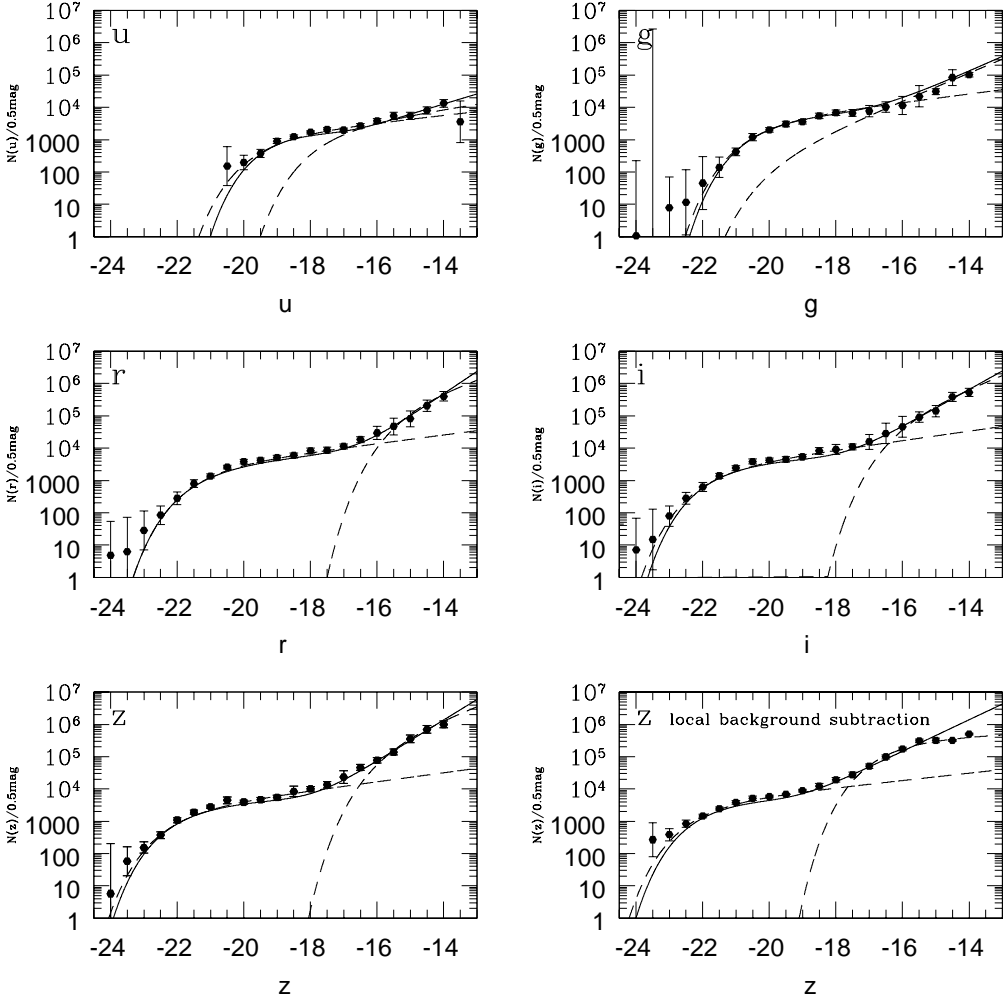}}
\end{minipage}
\end{center}
\caption{ 
The figure shows the Composite LF in the five Sloan bands calculated
within 1 Mpc $\rm{h}^{-1}$ aperture and with a global background
correction. For comparison we show also the composite LF in the z band
calculated with a local background subtraction. The solid line in each
plot is the result of the two Schechter components fit (2Scf), while
the dashed line are obtained with the single Schechter component fit
(SScf) at the bright and at the faint end of the LF. The 2Scf fit
perfectly reproduces the sum of the two single bright and faint
components.}
\label{lf}
\end{figure*}

\begin{figure*}
\begin{center}
\begin{minipage}{1.0\textwidth}
\resizebox{\hsize}{!}{\includegraphics{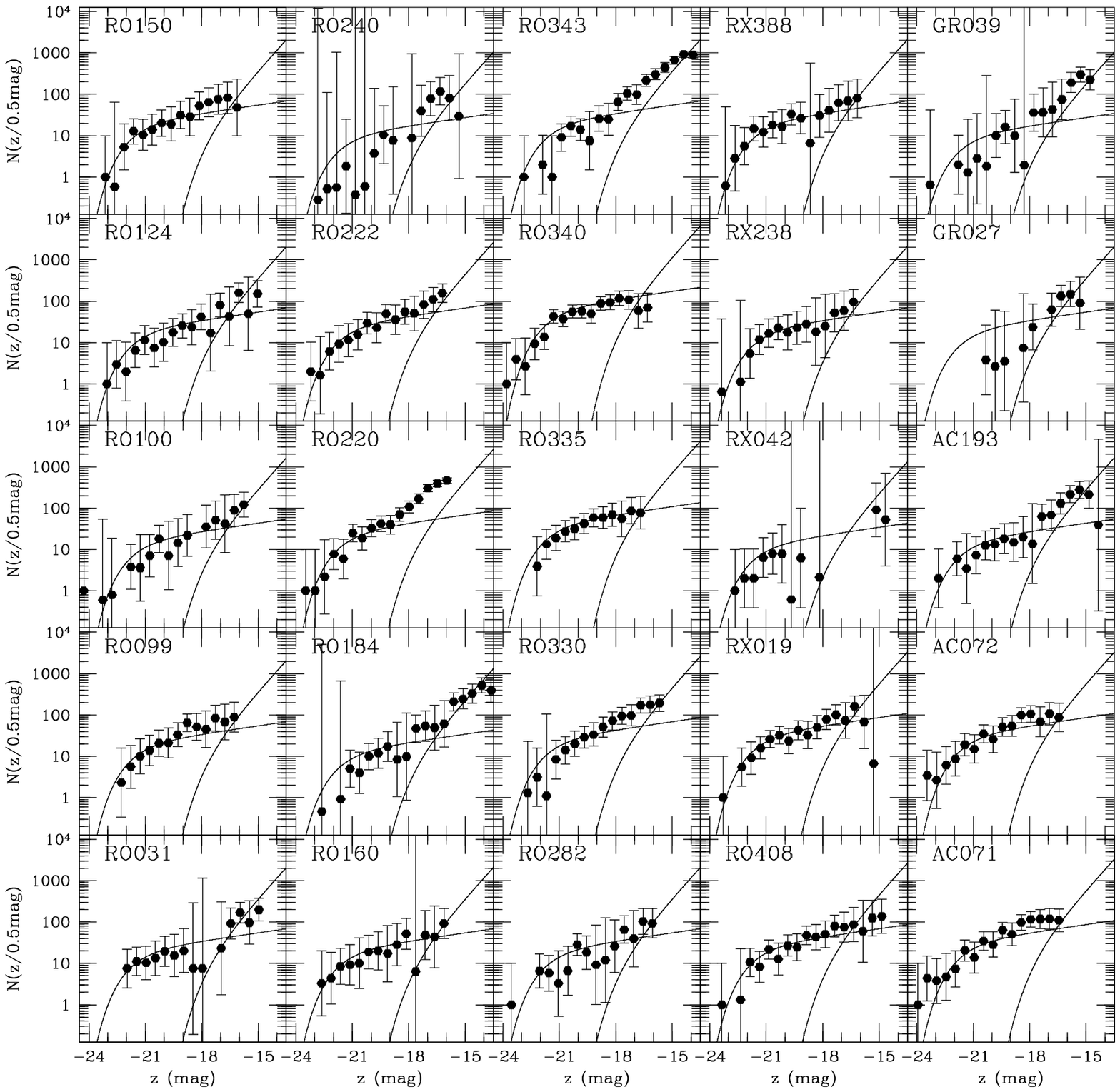}}
\end{minipage}
\end{center}
\caption{  
The plot shows the individual cluster luminosity functions for 25
clusters of the RASS-SDSS galaxy cluster catalog, calculated within 1
Mpc $\rm {h}^{-1}$ aperture and with a global background
subtraction. The solid line in the plots are the results of the SSfc
method applied to the corresponding Composite LF.The upturn of the
dwarf and the steepening of the LFs at the faint end is evident
in several clusters.}
\label{clus}
\end{figure*}

The formal error in the  $N_{cj}$ is computed according to:
\begin{equation}
\delta N_{cj}=\frac{N_{c0}}{m_j}[\sum_i{(\frac{\delta N_{ij}}{N_{i0}})^2}]^{(1/2)}
\end{equation}
where the $\delta N_{cj}$ and $\delta N_{ij}$ are the formal errors in
the $j\rm{th}$ bin of the Composite LF and of the $i\rm{th}$ cluster
LF. Since the $i\rm{th}$ cluster LF bin is given by the galaxy counts
corrected by the field subtraction, the formal error $\delta N_{ij}$
is calculated as the quadratic sum of the Poissonian error in the
counts and the background error.

It is easy to note that in the Colless (1989) prescriptions the
$j\rm{th}$ bin of the Composite LF represents just the mean fraction
of galaxies, with respect to the normalization region, of all the
clusters contributing to the $j\rm{th}$ bin.

The only restrictions in the use of the Colless method is that,
firstly, the magnitude limit of all the clusters has to be at least
fainter than the limit of the region of normalization ( $M < -19$ mag
in our case), secondly, the normalization region has to be large
enough to be representative of the richness of the cluster and,
finally, that the number of clusters contributing to each bin of
magnitude has to be statistically significant.  If these requirements
are satisfied, the Colless (1989) prescriptions can be used to build a
Composite LF which extends to the faintest magnitude limit of the
cluster sample, with an efficient use of the available data.
Therefore, we use the whole magnitude range available with our cluster
sample and we include in the Composite LF all the bins with at least
10 contributing clusters.

An alternative method has been recently proposed by Garilli et al. (1999),
whose prescriptions are:
\begin{equation}
N_{cj}=\frac{1}{m_j}\sum_i{N_{ij}w_i^{-1}}
\end{equation}
where, $N_{cj}$ and $N_{ij}$ have the same meaning as in the former
case, while $m_j$ is the number of clusters with limiting magnitude
deeper than the $j\rm{th}$ bin and $w_i$ is the weight of each
cluster, given by the ratio of the number of galaxies of the
$i\rm{th}$ cluster to the number of galaxies brighter than its
magnitude limit in all clusters with fainter magnitude limits (Stefano
Andreon private communication). The formal error in the Composite LF
is computed according to:
\begin{equation}
\delta N_{cj}=\frac{1}{m_j}\sqrt{\sum_i{N_{ij}w_i^{-2}}}.
\end{equation}

The important difference with the Colless (1989) prescriptions is that
in this case the Composite LF is not a simple mean of the galaxy
fraction in each bin (multiplied by a normalization constant), but a
weighted mean of the cluster galaxy number in each bin of magnitude.

\begin{figure*}
\begin{center}
\begin{minipage}{0.48\textwidth}
\resizebox{\hsize}{!}{\includegraphics{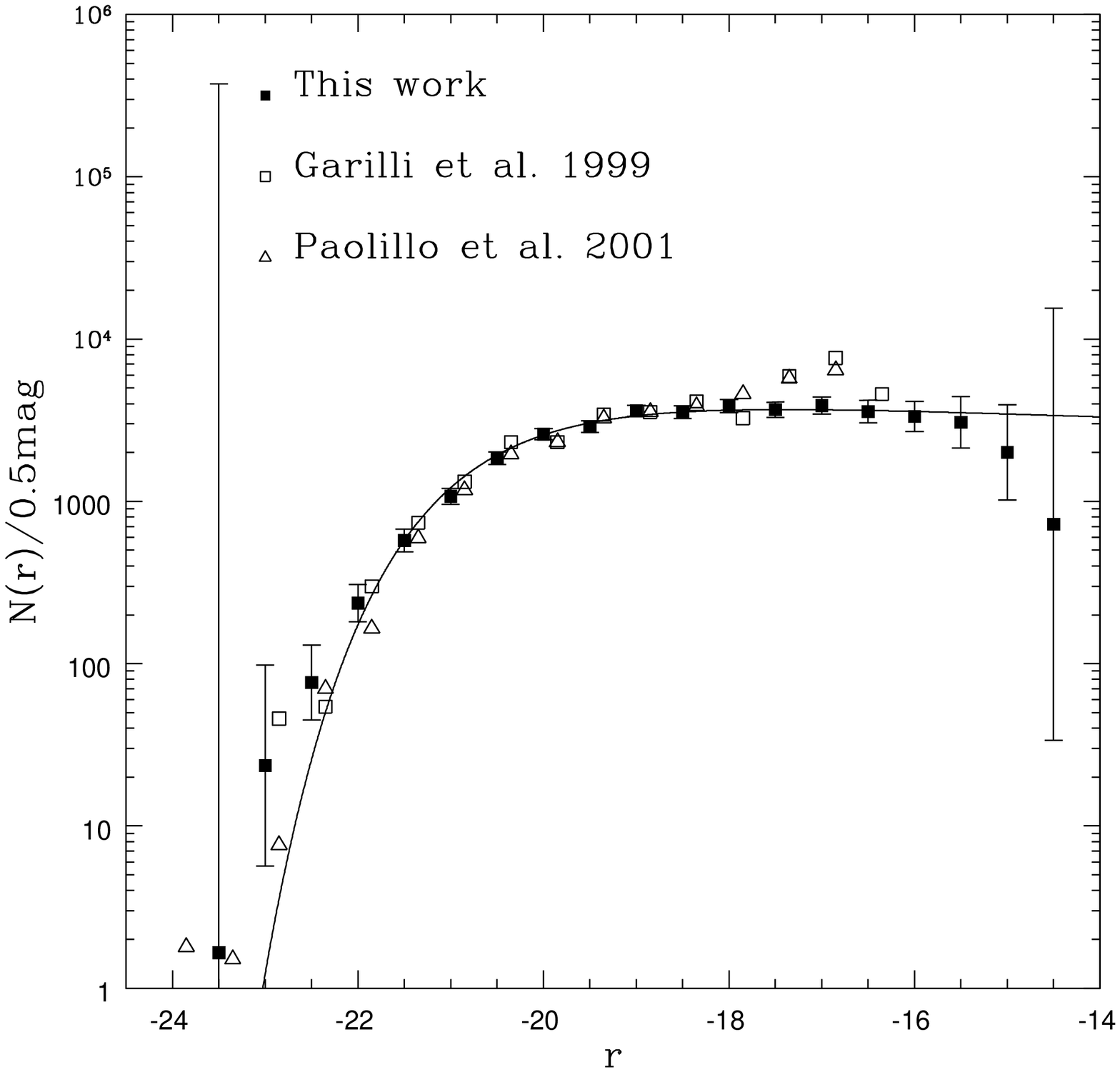}}
\end{minipage}
\begin{minipage}{0.48\textwidth}
\resizebox{\hsize}{!}{\includegraphics{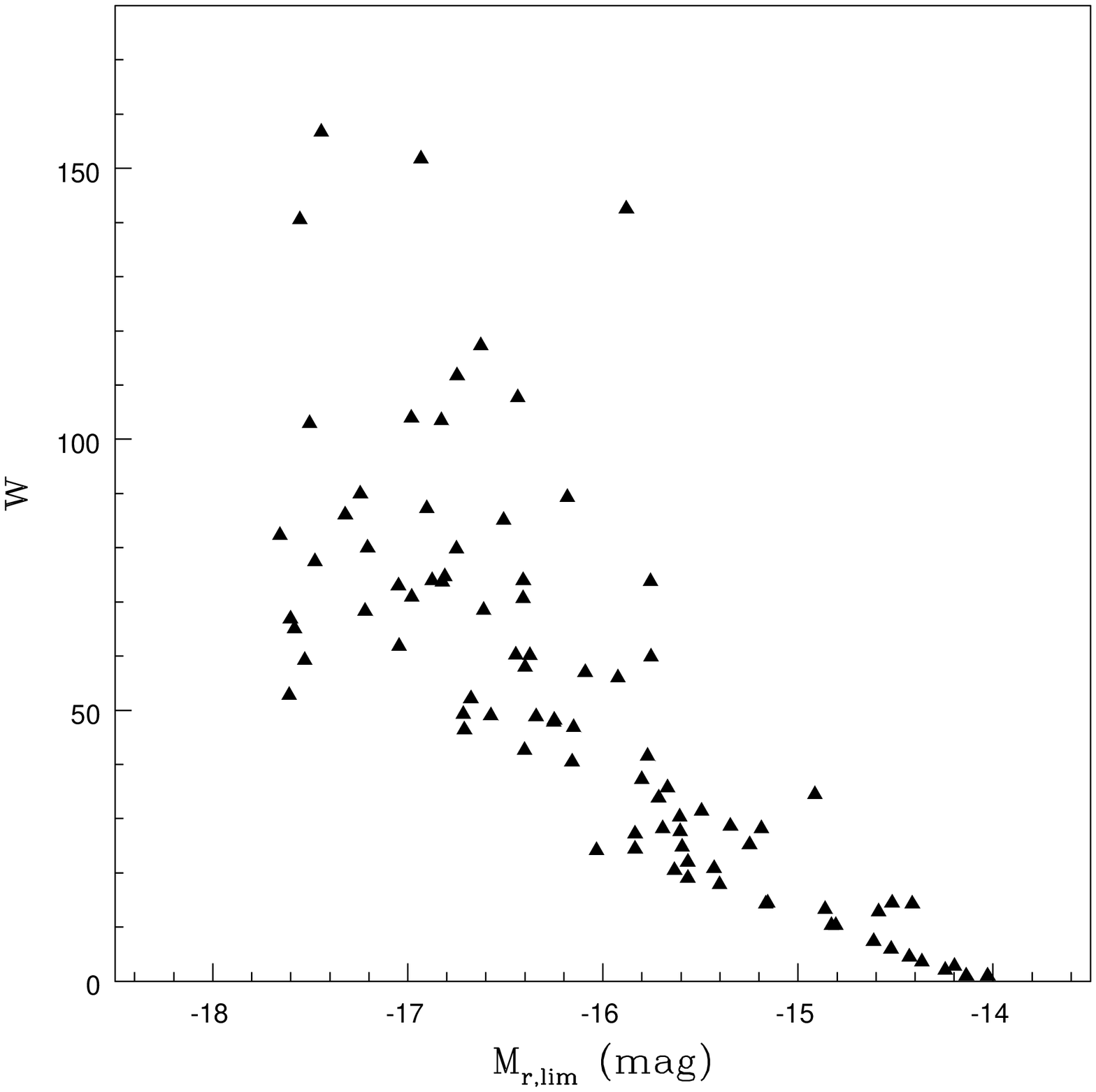}}
\end{minipage}
\end{center}
\caption{
The figure shows the results obtained applying the Garilli et
al. (1999) method. The plot on the left side shows the Composite LF in
the r Sloan band (filled squares). For comparison we plot also the
Composite LF obtained by Garilli et al. (1999) (empty squares) and by
Paolillo et al. (2001) (empty triangles). The three LFs agree very
well (within 1$\sigma$) in the characteristic magnitude and in the
faint end slope. However the Composite cluster LF obtained with this
prescription do not reproduce the main features observed in the
individual cluster LFs (see fig. \ref{clus}). The reason of the
disagreement is the weighing method in the Garilli's prescription. The
plot on the right side shows the dependence of the weight $w_i$ on the
magnitude limit of the single cluster. The system with very faint
$M_{lim}$, which contribute to the faint magnitude bins in the
composite LF, are heavily down-weighted. The bias explains the lack of
the upturn in the dwarf magnitude range observed in the individual
cluster LFs.}
\label{gar_pao}
\end{figure*}

\begin{figure}
\begin{center}
\begin{minipage}{0.5\textwidth}
\resizebox{\hsize}{!}{\includegraphics{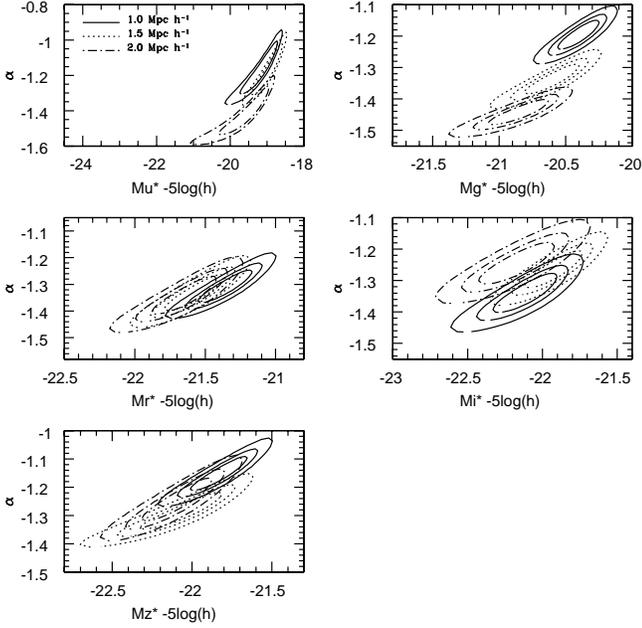}}
\end{minipage}
\end{center}
\caption{
1, 2 and  3$\sigma$ contours of  the best fit Schechter parameters for
the bright  component of the Composite  LF.   The contours are derived
with the  SScf method applied  to the bright  end of the Composite LF.
The LF  is calculated with  a global background subtraction  within 1,
1.5 and 2 Mpc $\rm {h}^{-1}$ apertures.}
\label{con}
\end{figure}

\begin{figure}
\begin{center}
\begin{minipage}{0.5\textwidth}
\resizebox{\hsize}{!}{\includegraphics{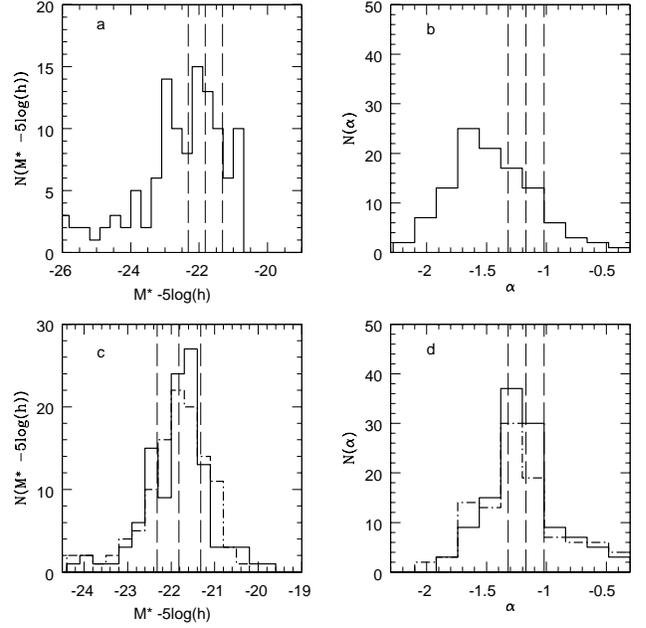}}
\end{minipage}
\end{center}
\caption{
Distribution of the Schechter parameters $\alpha$ and $M^*$ of the
individual cluster LFs in the sample, calculated in the z band within
1 Mpc $\rm{h}^{-1}$ and with a global background correction.  Plots
$a$ and $b$ show the distribution of $M^*$ $\alpha$, respectively,
obtained by fitting a single Schechter luminosity function to the
galaxies in the whole available magnitude range of each cluster.
Plots $c$ and $d$ show the distribution of $M^*$ and $\alpha$ ,
respectively, obtained by fitting a single Schechter luminosity
function to the galaxies brighter than the magnitude of the dwarf
upturn $M_Z \le -18$.  The dashed lines in plots $c$ and $d$ are the
distributions obtained from the fits with $\alpha $ and $M^*$ being
free parameters.  The solid line in plots $c$ is the
distribution of $M^*$ when $\alpha$ is fixed to the value of the
corresponding bright component of the Composite LF calculated with the
SScf method. The solid line in plot $d$ is the distribution of
$\alpha$ when $M^*$ is fixed to the value of the Composite LF.  The
vertical dashed lines in each plots indicate the value of the
corresponding parameters of the bright component in the Composite LF
and its 3$\sigma$ error interval.}
\label{histo}
\end{figure}

\begin{table*}
\caption{Schechter parameters of the Composite LF}
%\begin{center}
\begin{minipage}{1.0\textwidth}
\begin{sideways}
\begin{tabular}[b]{c|cccccccccc}
\hline
\multicolumn{1}{c}{ }& \multicolumn{2}{c}{u}& \multicolumn{2}{c}{g}& \multicolumn{2}{c}{r}& \multicolumn{2}{c}{i} & \multicolumn{2}{c}{z} \\ \hline
$r$ & $\alpha _u$ & $M*_u$ & $\alpha _g$ & $M*_g$ & $\alpha _r$ & $M*_r$ & $\alpha _i$ & $M*_i$ & $\alpha _z$ & $M*_z$  \\
\hline
\multicolumn{11}{c}{SScf - Bright  component, global background subtraction} \\ \hline
\hline
0.5 &  $ -1.31\pm 0.16$ &   $-19.59\pm 0.85$ &    $-1.18 \pm 0.05$ &  $-20.52\pm 0.26$ & $-1.29\pm 0.09$ &  $-21.54\pm 0.39$ & $-1.20\pm 0.06$ & $-21.77\pm 0.30$ & $-1.23 \pm 0.07$ & $-22.09\pm 0.30$ \\
1.0 &  $ -1.15\pm 0.15$ &   $-19.11\pm 0.48$ &    $-1.19 \pm 0.04$ &  $-20.39\pm 0.15$ & $-1.30\pm 0.06$ &  $-21.35\pm 0.19$ & $-1.07\pm 0.08$ & $-21.62\pm 0.15$ & $-1.16 \pm 0.06$ & $-21.86\pm 0.18$ \\
1.5 &  $ -1.16\pm 0.14$ &   $-18.92\pm 0.39$ &    $-1.33 \pm 0.04$ &  $-20.59\pm 0.20$ & $-1.33\pm 0.06$ &  $-21.57\pm 0.21$ & $-1.22\pm 0.06$ & $-21.76\pm 0.17$ & $-1.28 \pm 0.06$ & $-22.04\pm 0.25$ \\
2.0 &  $ -1.39\pm 0.13$ &   $-19.44\pm 0.61$ &    $-1.44 \pm 0.04$ &  $-20.83\pm 0.22$ & $-1.34\pm 0.07$ &  $-21.63\pm 0.22$ & $-1.25\pm 0.06$ & $-22.19\pm 0.25$ & $-1.25 \pm 0.07$ & $-22.11\pm 0.22$ \\
\hline								   							   

\multicolumn{11}{c}{SScf - Bright  component, local background subtraction} \\ \hline
\hline
0.5 &  $ -1.28\pm 0.15$ &   $-19.38\pm 0.63$ &    $-1.25 \pm 0.04$ &  $-20.64\pm 0.23$ & $-1.41\pm 0.07$ &  $-21.81\pm 0.43$ & $-1.33\pm 0.05$ & $-22.13\pm 0.33$ & $-1.33 \pm 0.05$ & $-22.44\pm 0.27$ \\
1.0 &  $ -1.34\pm 0.08$ &   $-18.93\pm 0.18$ &    $-1.44 \pm 0.05$ &  $-20.76\pm 0.19$ & $-1.33\pm 0.06$ &  $-21.40\pm 0.20$ & $-1.25\pm 0.06$ & $-21.63\pm 0.16$ & $-1.28 \pm 0.06$ & $-21.99\pm 0.18$ \\
1.5 &  $ -1.37\pm 0.17$ &   $-19.30\pm 1.03$ &    $-1.24 \pm 0.10$ &  $-20.49\pm 0.19$ & $-1.40\pm 0.05$ &  $-21.71\pm 0.19$ & $-1.47\pm 0.04$ & $-22.14\pm 0.18$ & $-1.35 \pm 0.06$ & $-22.07\pm 0.16$ \\
2.0 &  $ -1.38\pm 0.10$ &   $-19.40\pm 0.71$ &    $-1.02 \pm 0.16$ &  $-20.35\pm 0.21$ & $-1.51\pm 0.06$ &  $-21.93\pm 0.24$ & $-1.54\pm 0.03$ & $-22.31\pm 0.16$ & $-1.51 \pm 0.05$ & $-22.46\pm 0.19$ \\
\hline

\multicolumn{6}{c|}{SScf - Faint  component, global background subtraction } & \multicolumn{5}{c}{SScf - Faint component, local background subtraction }\\ \hline
\multicolumn{1}{c}{$r$} & \multicolumn{1}{c}{$\alpha _u$} & \multicolumn{1}{c}{$\alpha _g$} & \multicolumn{1}{c}{$\alpha _r$} & \multicolumn{1}{c}{$\alpha _i$} & \multicolumn{1}{c|}{$\alpha _z$} & \multicolumn{1}{c}{$\alpha _u$} & \multicolumn{1}{c}{$\alpha _g$} & \multicolumn{1}{c}{$\alpha _r$} & \multicolumn{1}{c}{$\alpha _i$} & \multicolumn{1}{c}{$\alpha _z$}  \\
\hline
0.5 &  $ -1.50\pm 0.35$ &  $-1.98\pm 0.38$ & $-1.96\pm 0.24$ & $-1.81\pm 0.15$  & \multicolumn{1}{c|}{$-1.80\pm 0.16$} &  $ -1.60\pm 0.25$ &  $-2.16\pm 0.09$ & $-2.18\pm 0.04$ & $-1.98\pm 0.08$  & $-2.18\pm 0.03$ \\
1.0 &  $ -1.40\pm 0.14$ &  $-1.88\pm 0.24$ & $-1.54\pm 0.50$ & $-1.61\pm 0.08$  & \multicolumn{1}{c|}{$-2.24\pm 0.10$} &  $ -1.50\pm 0.33$ &  $-2.45\pm 0.47$ & $-1.83\pm 0.06$ & $-2.27\pm 0.07$  & $-1.72\pm 0.05$ \\ 
1.5 &  $ -1.69\pm 0.07$ &  $-1.73\pm 0.25$ & $-2.11\pm 0.37$ & $-1.74\pm 0.21$  & \multicolumn{1}{c|}{$-2.27\pm 0.05$} &  $ -1.73\pm 0.03$ &  $-1.73\pm 0.05$ & $-1.53\pm 0.07$ & $-1.90\pm 0.07$  & $-1.78\pm 0.05$ \\
2.0 &  $ -1.53\pm 0.06$ &  $-2.05\pm 0.20$ & $-1.74\pm 0.11$ & $-2.09\pm 0.11$  & \multicolumn{1}{c|}{$-2.44\pm 0.11$} &  $ -1.66\pm 0.03$ &  $-1.64\pm 0.13$ & $-2.08\pm 0.06$ & $-1.91\pm 0.05$  & $-2.35\pm 0.02$ \\
\hline

\multicolumn{11}{c}{2Scf - Bright component, global background subtraction } \\ \hline
\hline
0.5 & $ -0.92\pm  0.13 $  & $ -18.00\pm  0.50  $ &  $ -1.30\pm  0.13 $  &  $-20.75\pm  0.46 $  &  $ -1.30\pm  0.12 $ &  $ -21.50\pm  0.51  $ &  $ -1.09\pm  0.13  $ & $ -21.54\pm  0.41 $  &  $ -1.23\pm  0.11 $  & $ -22.19\pm  0.44 $\\
1.0 & $ -1.59\pm  0.13 $  & $ -19.24\pm  0.53  $ &  $ -0.55\pm  0.17 $  &  $-19.67\pm  0.20 $  &  $ -1.03\pm  0.13 $ &  $ -20.90\pm  0.26  $ &  $ -1.14\pm  0.11  $ & $ -21.56\pm  0.26 $  &  $ -1.07\pm  0.12 $  & $ -21.73\pm  0.27 $\\	
1.5 & $ -1.27\pm  0.22 $  & $ -19.40\pm  0.23  $ &  $ -1.41\pm  0.16 $  &  $-20.80\pm  0.46 $  &  $ -1.39\pm  0.07 $ &  $ -21.50\pm  0.23  $ &  $ -1.20\pm  0.04  $ & $ -21.98\pm  0.26 $  &  $ -1.06\pm  0.16 $  & $ -21.69\pm  0.38 $\\
2.0 & $ -1.50\pm  0.17 $  & $ -20.59\pm  0.09  $ &  $ -1.58\pm  0.15 $  &  $-21.53\pm  0.82 $  &  $ -1.06\pm  0.10 $ &  $ -21.24\pm  0.37  $ &  $ -0.94\pm  0.22  $ & $ -21.61\pm  0.49 $  &  $ -1.29\pm  0.03 $  & $ -22.17\pm  0.36 $\\
\hline
\multicolumn{11}{c}{2Scf - Bright  component, local background subtraction } \\ \hline
\hline
0.5 & $ -0.68\pm  0.18 $  & $ -18.32\pm  0.33  $ &  $ -1.24\pm  0.16 $  &  $-20.62\pm  0.46 $  &  $ -1.23\pm  0.19 $ &  $ -21.36\pm  0.60  $ &  $ -1.16\pm  0.19  $ & $ -21.79\pm  0.62 $  &  $ -1.22\pm  0.12 $  & $ -22.14\pm  0.44 $\\
1.0 & $ -0.95\pm  0.23 $  & $ -19.53\pm  0.30  $ &  $ -1.23\pm  0.11 $  &  $-20.39\pm  0.27 $  &  $ -1.05\pm  0.13 $ &  $ -20.95\pm  0.27  $ &  $ -1.17\pm  0.13  $ & $ -21.64\pm  0.29 $  &  $ -1.06\pm  0.12 $  & $ -21.70\pm  0.26 $\\
1.5 & $ -1.71\pm  0.13 $  & $ -20.36\pm  0.26  $ &  $ -0.91\pm  0.28 $  &  $-20.23\pm  0.34 $  &  $ -0.76\pm  0.13 $ &  $ -20.86\pm  0.20  $ &  $ -1.11\pm  0.09  $ & $ -21.51\pm  0.21 $  &  $ -1.02\pm  0.12 $  & $ -21.71\pm  0.21 $\\
2.0 & $ -0.96\pm  0.49 $  & $ -18.75\pm  0.76  $ &  $ -0.99\pm  0.23 $  &  $-20.15\pm  0.35 $  &  $ -1.03\pm  0.14 $ &  $ -21.19\pm  0.23  $ &  $ -1.27\pm  0.11  $ & $ -21.82\pm  0.26 $  &  $ -1.46\pm  0.06 $  & $ -22.41\pm  0.26 $\\
\hline
\multicolumn{11}{c}{2Scf - Faint  component, global background subtraction } \\ \hline
\hline
0.5 & $ -0.88\pm  0.18 $  & $ -18.92\pm  0.45  $ &  $ -2.44\pm  0.25 $  &  $-16.99\pm  0.31 $  &  $ -2.38\pm  0.15 $ &  $ -17.76\pm  0.23  $ &  $ -2.09\pm  0.07  $ & $ -18.34\pm  0.19 $  &  $ -2.28\pm  0.08 $  & $ -18.59\pm  0.21 $\\
1.0 & $  0.00\pm  0.00 $  & $ -18.09\pm  0.37  $ &  $ -2.04\pm  0.03 $  &  $-17.89\pm  0.11 $  &  $ -2.01\pm  0.05 $ &  $ -18.40\pm  0.15  $ &  $ -2.36\pm  0.05  $ & $ -18.86\pm  0.16 $  &  $ -2.22\pm  0.06 $  & $ -19.09\pm  0.17 $\\
1.5 & $ -2.65\pm  0.90 $  & $ -15.43\pm  0.44  $ &  $ -2.54\pm  0.18 $  &  $-17.18\pm  0.25 $  &  $ -2.79\pm  0.14 $ &  $ -17.40\pm  0.18  $ &  $ -2.83\pm  0.07  $ & $ -18.00\pm  0.14 $  &  $ -2.70\pm  0.07 $  & $ -18.71\pm  0.12 $\\
2.0 & $  0.00\pm  0.05 $  & $ -17.31\pm  0.01  $ &  $ -2.52\pm  0.30 $  &  $-17.52\pm  0.57 $  &  $ -2.03\pm  0.08 $ &  $ -18.63\pm  0.21  $ &  $ -2.21\pm  0.08  $ & $ -18.91\pm  0.23 $  &  $ -2.76\pm  0.05 $  & $ -18.64\pm  0.14 $\\
\hline
\multicolumn{11}{c}{2Scf - Faint  component,local background subtraction } \\ \hline
\hline
0.5 & $  0.00\pm  0.00 $  & $ -15.63\pm  0.26  $ &  $ -2.23\pm  0.25 $  &  $-17.26\pm  0.32 $  &  $ -2.18\pm  0.18 $ &  $ -18.12\pm  0.26  $ &  $ -2.17\pm  0.19  $ & $ -18.51\pm  0.22 $  &  $ -2.34\pm  0.12 $  & $ -18.57\pm  0.20 $\\
1.0 & $ -1.64\pm  0.25 $  & $ -19.79\pm  4.04  $ &  $ -2.84\pm  0.13 $  &  $-17.27\pm  0.13 $  &  $ -2.02\pm  0.05 $ &  $ -18.42\pm  0.15  $ &  $ -2.45\pm  0.13  $ & $ -18.96\pm  0.17 $  &  $ -2.21\pm  0.06 $  & $ -19.08\pm  0.16 $\\
1.5 & $ -0.76\pm  5.46 $  & $ -18.71\pm  0.87  $ &  $ -1.86\pm  0.07 $  &  $-18.49\pm  0.26 $  &  $ -1.92\pm  0.03 $ &  $ -18.94\pm  0.13  $ &  $ -2.13\pm  0.03  $ & $ -19.07\pm  0.13 $  &  $ -2.25\pm  0.05 $  & $ -19.45\pm  0.13 $\\
2.0 & $ -1.62\pm  0.04 $  & $ -18.71\pm  2.17  $ &  $ -2.26\pm  0.07 $  &  $-18.26\pm  0.18 $  &  $ -2.22\pm  0.04 $ &  $ -19.06\pm  0.15  $ &  $ -2.54\pm  0.10  $ & $ -19.21\pm  0.14 $  &  $ -2.61\pm  0.06 $  & $ -19.16\pm  0.12 $\\
\hline	

\end{tabular}
\end{sideways}	
\begin{sideways}
\parbox{24cm}{
The Table lists the Schechter parameters of the bright and the faint
end of the composite LF.  The results are obtained with a single
Schechter component fit (SScf) and with a two Schechter components fit
(2Scf).  For each case the fit procedure was applied to the composite
LF calculated within 4 different clustercentric distances, 0.5, 1.0,
1.5 and 2.0 Mpc $\rm{h}^{-1}$ and with different background
corrections.}
\label{res}
\end{sideways}	
\end{minipage}
%\end{center}
\end{table*}

\section {Results}
Fig.  \ref{lf} shows the Composite LF obtained with the Colless (1989)
prescription with a global and local background corrections.  In both
cases, the Composite LF shows a clear bimodal behavior, showing the
upturn of the dwarf galaxies in the magnitude region $-18 \le M \le
-16 $, depending on the waveband.  We apply two different approaches
in fitting the Composite LF.  We divide the Composite LF in two
components, a ``brigh-end'' and a ``faint-end'' Composite LF, locating
by eye the upturn of the dwarf galaxies.  We then fit the two
components separately using a Single Schechter component fit (SScf).
As a second approach, we fit the whole available range of magnitude
of the Composite LF with the sum of the two Schechter components ( 2
Schechter components fit, 2Scf).  The dashed lines in fig.  \ref{lf}
are the results of the SScf method, while the solid line is the fit
resulting from the 2Scf procedure.  There is a very good agreement
between the results of the methods applied. Table \ref{res} lists the
values of the Schechter parameters of the bright and the faint LF
components obtained with different fitting procedures.  The Composite
LFs are calculated using different background subtractions and within
different cluster radii (from 0.5 to 2.0 Mpc
$\rm{h}^{-1}$). Fig. \ref{clus} shows the individual cluster LFs for a
subsample of 25 systems of the RASS-SDSS galaxy cluster catalog.  We
overplotted the results of the SScf method applied to the
corresponding Composite LF.

For comparison we also applied the method proposed by Garilli et
al. (1999). The plot on the feft side in fig. \ref{gar_pao} shows the
results obtained applying that method.  It is clear that the upturn in
the faint magnitude region disappears completely and the composite LF
is well fitted by a single Schechter function.  The results obtained
with the Garilli et al. (1999) prescription do not agree within the
errors with the results obtained with the Colless method, and show a
much flatter LF with a fainter $M^*$ in all the wavebands. Instead,
there is a very good agreement ( 1 $\sigma$) with the Schechter
parameters obtained by Galilli et al.  (1999) and Paolillo et al.
(2001), which applied the same method to derive the composite LF.  The
Composite LF obtained with this prescription is not a good
representation of the mean cluster LF since it does not reproduce the
features visible in the individual cluster LFs (see fig. \ref{clus}).
The reason of the disagreement between the Composite LF obtained with
the Garilli's method and the individual LFs is due to the different
weighting method applied by Garilli et al. (1999). As shown in the
plot on the right side of fig. \ref{gar_pao}, the weight in the
Garilli et al. (1999) method depends strongly on the cluster magnitude
limits. The weight $w_i$ is a decreasing function of the cluster
$M_{lim}$, therefore, the clusters with fainter $M_{lim}$, which
contribute to the faint magnitude bins, are heavily
down-weighted. This bias explains the lack of the dwarf upturn in the
Composite LF. In conclusion, the Garilli's method provides a biased
estimation of the Composite cluster LF.

In the following analysis we consider only the results obtained with
the Colless (1989) prescription.

\subsection{The bright end}
The Schechter parameters $\alpha$ and $M^*$ obtained for the Composite
LF derived with the Colless (1989) prescriptions with the local and
global background corrections agree very well at any radius (in the
worse cases within 1.5 $\sigma$).  There is also a very good agreement
within the errors for $M^*$ obtained with different fitting
procedures.  The slope of the bright component calculated with
the SScf method is systematically steeper than the slope obtained with
the 2Scf procedure.  This is due to the fact that in the 2Scf method
the fitting function is the sum of two components. Consequently, the
slope of the bright component does not represent only the bright
galaxies population but depends also on the slope and $M^*$ of the
second (faint) component.  Therefore, in the following analysis we
consider the parameters of the bright component obtained with the SScf
method as representative of the bright galaxies population in
clusters.

Fig. \ref{con} shows the error contours of these Schechter parameters
calculated for the Composite LFs measured within 3 different cluster
radii, 1.0, 1.5 and 2.0 Mpc $\rm{h}^{-1}$ in the case of a global
background correction.  $\alpha$ and $M^*$ seem not to depend on the
clustercentric distance since the error contours overlap in any
wavebands except for the g band.  However, the behavior of the
Schechter parameters in this band is not confirmed by the same
Composite LF calculated with the local background
subtraction. Therefore, we can conclude that there is no significant
difference in the bright end LF measured in different aperture radii.

\begin{figure}
\begin{center}
\begin{minipage}{0.5\textwidth}
\resizebox{\hsize}{!}{\includegraphics{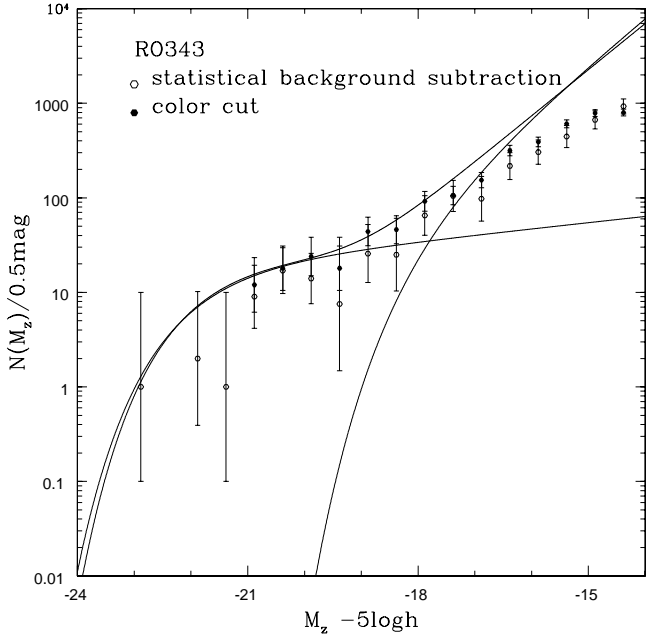}}
\end{minipage}
\end{center}
\caption{ 
The plot shows two different methods of background subtraction.  The
cluster RO343 is one of the clusters showing a significant steepening
of the LF in the faint magnitude range. The filled points indicate the
cluster LF obtained with a cut in the $g-r$ - $r-i$ plane (Garilli et
al. 1999).  The empty points are the LF obtained with the statistical
local background correction applied to obtain the composite LF
analysed in the paper.  The methods of background subtraction agree
perfectly within the errors.  The error bars in the color-cut method
are the Poissonian error in the galaxy counts. The color cut method
excludes all the galaxies redder than the color of an elliptical
galaxy at the cluster redshift. Therefore, the steepening in the faint
end can not be due to galaxies at higher redshift in a second cluster
or in the large scale structure behind the cluster, but should be due
to the presence of a real cluster population. A bluer color cut deletes
the contribution of the bright elliptical cluster galaxies leaving the
faint end LF.  This implies that the faint end LF is dominated by
late-type galaxies. }
\label{color}
\end{figure}

To check the universality of the cluster LF, we compare the Schechter
parameters of the bright component of Composite LF with the Schechter
parameters derived by fitting the individual cluster LFs.  Fig.
\ref{histo}  shows the   distributions of $M^*$    and $\alpha$ of  the
individual cluster LFs derived in the z band within 1 Mpc
$\rm{h}^{-1}$ from the cluster center and with a global background
correction.  The vertical dashed lines in the plots show the value of
the corresponding Composite LF parameter and the 3$\sigma$ error
interval.  The plots $a$ and $b$ in the fig.  \ref{histo} show the
distributions of $M^*$ and $\alpha$ when a single Schechter luminosity
function is fitted to the galaxies in the whole available magnitude
range of each cluster (including the dwarf region).  It is clear from
those distributions that the ``bright end'' Composite LF is not a good
representation of the mean behavior of the individual LFs: the
individual LFs seems to be systematically steeper and the dispersion
of $M^*$ is bigger than 2 magnitudes.  The distributions of both
parameters change drastically if the galaxies in the dwarfs region are
excluded from the fits.  Plots $c$ and $d$ of the fig.\ref{histo}
clearly show that the distributions become in both cases close to a
Gaussian with the maximum coincident with the value of the
corresponding Composite LF parameter.  The dispersion of the
distribution of $\alpha$ seems to be bigger than the 3$\sigma$ error
interval of the Composite LF parameter. Therefore, we could conclude
that the Composite LF is a very good representation of the mean
behavior of the individual cluster ``bright end'' LFs, but it is not
universal. Nevertheless, it is important to stress that we assume
that the dwarf upturn of all the clusters in the sample has the same
location observed in the Composite LF ($M_z \sim -18$).  A brighter
upturn could give a steeper individual LF and, therefore, it could
explain the excess of clusters in the region $\alpha \le -1.3$ in the
plot $d$ of fig. \ref{histo}.

\begin{figure}
\begin{center}
\begin{minipage}{0.5\textwidth}
\resizebox{\hsize}{!}{\includegraphics{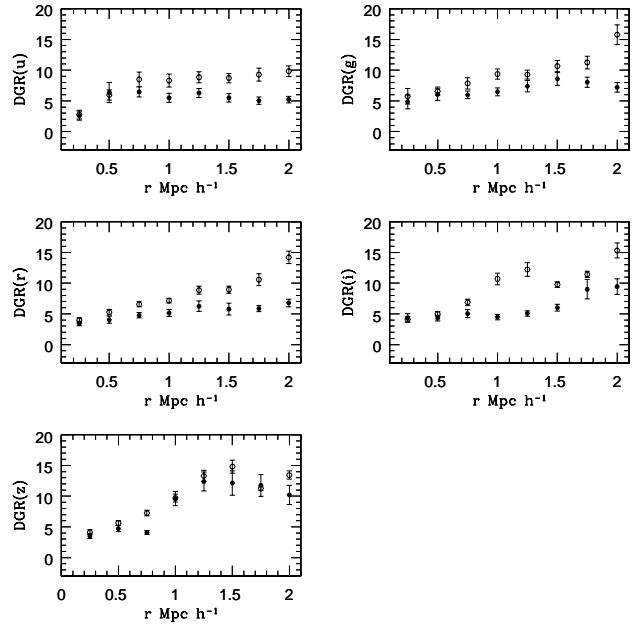}}
\end{minipage}
\end{center}
\caption{
Dwarf to Giant Ratio (DGR) as a function of the cluster radii in the 5
wavebands. DGR is derived from the Composite LF calculated with a
global background correction (filled points) and a local background
correction (empty points).}
\label{dgr}
\end{figure}

\begin{figure}
\begin{center}
\begin{minipage}{0.5\textwidth}
\resizebox{\hsize}{!}{\includegraphics{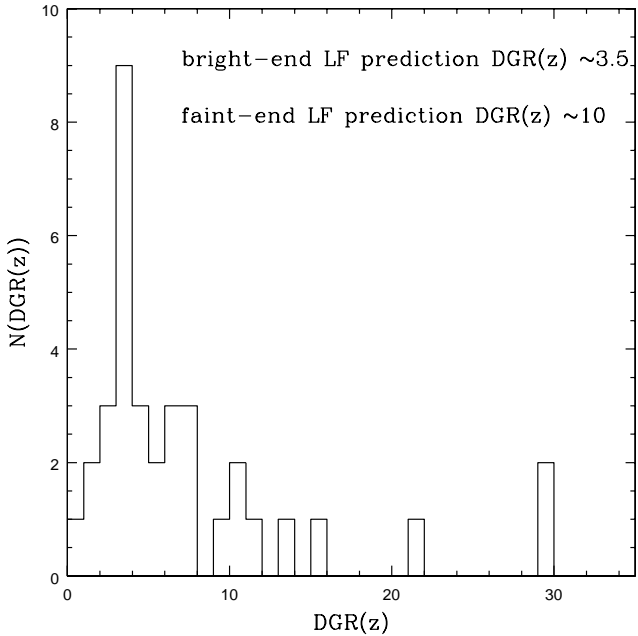}}
\end{minipage}
\end{center}
\caption{ 
Distribution of the DGR(z). DGR(z) is defined as as the ratio between
the number of galaxies brighter than -20 mag and the number of
galaxies in the magnitude range $-18. \le M_z \le -16.5$.  The faint
magnitude range in the definition of DGR(z) is large enough to be
representative of the dwarf population while the number of clusters
with magnitude limits fainter than -16.5 mag is still large (35
systems) to be statistically significant.  The value of DGR(z)
predicted by the ``bright end'' Composite LF (without the dwarf
population) is 3.5, while the value predicted with the presence of the
``faint-end'' Composite LF if around 10.}
\label{gdr_histo}
\end{figure}

\subsection{The faint end}
The results obtained for the fits of the faint LF components with
different fitting procedures, background corrections and cluster
apertures are listed in Table \ref{res}.  For the SScf method we
report only the slope of the faint-end component in each band and
not the values of $M^*$.  In fact, the faint end of the composite LF
does not contain a sufficient number of points to constrain in a
meaningful way the characteristic magnitude, and the statistical
errors of $M^*$ are larger than 1 mag.  We listed in the same table
$\alpha$ and $M^*$ measured with the 2Scf method.  In this case the
characteristic magnitude of the faint end is constrained by the slope
of the bright component.

As Table 3 shows, the ``faint end'' Composite LF is much more steeper
than the ``bright end'' LF at any radius and in any passband with both
the fitting procedures. There is a discrepancy between the values of
the slope of the SScf and the 2Scf methods in all the analysed
cases. The reason of the disagreement is the same observed for the
slopes of the bright component. The mean value of $\alpha$ derived
with the SScf method in the case of a global background correction is
1.60 in u, 1.84 in g, 1.81 in r, 1.76 in i and 2.07 in z.  The slope
do not show a dependence on the waveband and on the distance from the
cluster center. The result is confirmed also by the values given by
the 2Scf procedure.

Valotto et al.  (2001) use a numerical simulation of a hierarchical
universe to show that many ``clusters'' identified from two
dimensional galaxy distributions might result principally from the
projection of a large-scale structure alone the line of sight.  They
suggest that attempts to derive the LF for these ``clusters'' using
the standard background subtraction procedure lead to derive a LF with
a steep faint-end slope, despite the fact that the actual input LF had
a flat faint-end.  Since the RASS-SDSS galaxy cluster sample comprises
only clusters detected in X-rays, all the systems contributing to the
``fain-end'' Composite LF (26 clusters with $M_{zlim} \ge -16$) are
not a projection effect but are real clusters.  Moreover, the use of a
local background subtraction, which takes into account the presence of
large-scale structure, confirms the steepening of the Composite LF
observed with the global background subtraction. Valotto et al. (2004)
compare the composite luminosity function of a optically selected
sample of clusters with an X-ray selected sample of systems from the
RASS1 bright clusters catalog of De Grandi et al. (1999). They show
that the composite LF of the former sample presents a steep faint end
due to projection effects, while the composite LF of the X-ray
selected sample is flat with a slope of -1.1 in the magnitude range
$M_{b_J} \le -16.5$. Our results are still in agreement with Valotto
et al. (2004), since we are observing a much more faint population
with $ -16 \le M_g -5log(h) \le -14$.  Nevertheless, one could still
suspect that the observed steepening of the faint end in the
individual clusters is due to a second object at higher redshift and
on the same line of sight.  In fact, in this case, both the global and
local background corrections would not substract this contribution.
To test this possibility, we use the SDSS spectroscopic redshifts to
check the presence of galaxy overdensities at higher redshift and in
the same line of sight of the systems of interest.  Only RO184 shows a
second object in background while all the others clusters with and
without steepening in the individual LF present the same single peak
redshift distribution.

As additional test we try to measure the individual cluster LF with a
color cut method in the same way of Garilli et al.  (1999).  We use
the $g-r$ and $r-i$ galaxy colors defined in Fukugita et al. (1995).
We define our color cut in order to exclude all the galaxies redder
than the expected color of the ellipticals at the cluster redshift,
and the late type galaxies in foreground.  We observe that the systems
with a significant steepening in the individual LF obtained with
statistical background subtraction show the same feature also with the
color cut method (fig. \ref{color}). This implies that we are not
observing the contribution of large scale structures but a real
cluster faint population.  Moreover, we observe that the faint-end of
those clusters is due to galaxies with colors compatible with spiral
galaxies at the redshift of the cluster.  Finally, we can conclude
that the observed steepening of the Composite LF in the considered
magnitude range is real.

It is important to stress that, even if the Schechter function with
the values reported in Table \ref{res} offers a very good fit to the
data ( reduced $\chi ^2 \le 1.5$ in the worst case), the ``faint end''
Composite LF contains only few points.  Therefore, the slope $\alpha$
has to be considered as a good indicator of the steepening of the LF in
this magnitude region, but does not allow to a detailed analysis of
the behavior of the Composite LF. To study in more detail the behavior
of the ``fain-end'' Composite LF as a function of the waveband and of
the distance from the center, we define the Dwarf to Giant Ratio
(DGR) in each band as the ratio between the number of galaxies of the
``faint end'' Composite LF to the number of galaxies of the ``bright
end'' Composite LF.  We define DGR as the ratio between the number of
galaxies in the magnitude range $ -18 \le M \le -16.5$ and the number
of galaxies brither than -20 mag (except in the u band where we count
the galaxies brighter than -19 mag).  Fig.  \ref{dgr} shows the
behavior of DGR in each band as a function of the clustercentric
distance. The filled points are derived in the case of a global
background subtraction, while the empty points in the case of a local
background subtraction. The two results do not agree perfectly on the
DGR value, but they reproduce that same dependence on the
clustercentric distance.  In each waveband the DGR seems to slightly
increase from the very center, 0.3 Mpc $\rm{h}^{-1}$, to 1.0 Mpc
$\rm{h}^{-1}$. The mean value of DGR increases from the u band (5) to
the z band (10).

\begin{figure}
\begin{center}
\begin{minipage}{0.5\textwidth}
\resizebox{\hsize}{!}{\includegraphics{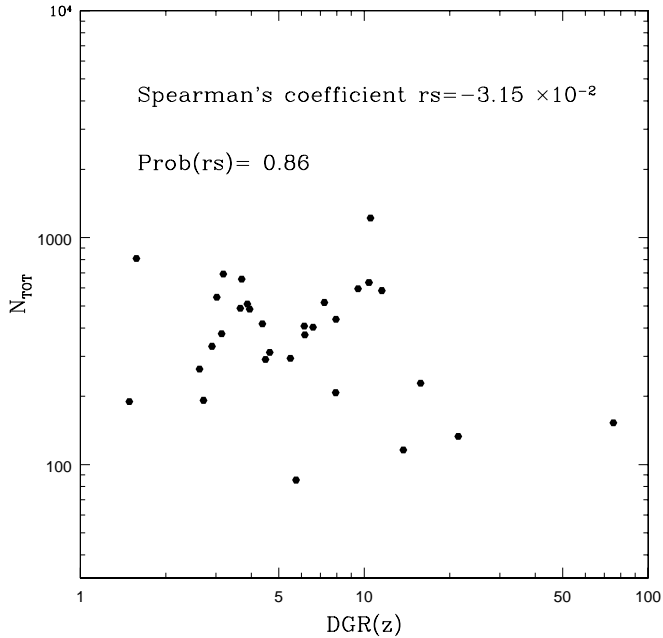}}
\end{minipage}
\end{center}
\caption{ 
DGR    versus the cluster  richness   in   the z  band   within 1  Mpc
$\rm{h}^{-1}$ in the case of global background subtraction. }
\label{gdr1}
\end{figure}

\begin{figure}
\begin{center}
\begin{minipage}{0.5\textwidth}
\resizebox{\hsize}{!}{\includegraphics{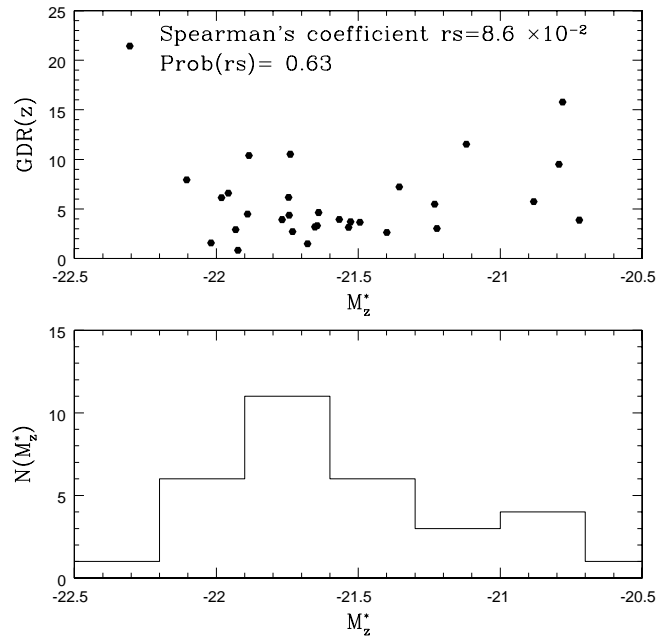}}
\end{minipage}
\end{center}
\caption{ 
The upper panel shows DGR versus the characteristic magnitude of the
individual bright component LF derived within 1 Mpc $\rm{h}^{-1}$ in
the case of global background subtraction. The bottom panel shows the
histogram of $M^*$ for the subsample of clusters. The histogram mimics
the behavior of the whole sample showed in the panel c) of
fig. \ref{histo}}
\label{gdr2}
\end{figure}

To test whether the ``faint-end'' Composite LF is a standard
representation of the dwarf population of galaxy clusters or if it is
due to the contribution of few particular clusters, we define a DGR
for the individual objects and compare it to the DGR of the Composite
LF.  Fig. \ref{gdr_histo} shows the distribution of the DGR calculated
for the single clusters in the z band.  The faint magnitude range in
the definition of DGR(z) is large enough to be representative of the
dwarf population while the number of clusters with magnitude limits
fainter than -16.5 mag is still large (35 systems) to be statistically
significant.  It is important to stress that the value of DGR(z)
predicted by the ``bright end'' Composite LF (without the dwarf
population) is 3.5, while the value predicted with the presence of the
``faint-end'' Composite LF is around 10.  As shown in fig.
\ref{gdr_histo}, there is   a large  spread  in  the distribution   of
DGR(z). The histogram in the figure shows a clear peak around the
value predicted by the ``bright end'' LF (3.5), and a large number of
objects (1/2) at values larger than this. This result indicates that
the behavior of the faint end LF in not universal. We conclude that
there seem to exist two different kind of cluster populations
depending on the excess of the dwarf galaxies.

To study the spread of the distribution in DGR(z), we plot DGR(z)
versus the cluster richness and versus the $M^*$ of the individual
cluster brigt-end LF, as shown in figs.  \ref{gdr1} and \ref{gdr2},
respectively.  We do not find any correlation between the
parameters. The Spearman's rank coefficient is very low in both cases
and with a probability of non correlation close to 1.  To further
understand the nature of the observed spread in the behavior of the
faint galaxy population in clusters, those systems should be well
studied individually.

It is straightforward to notice that in our analysis we do not take
into account possible low surface brightness selection
effects. Unfortunately, the analysis of the completness limits in
surface brightness of the SDSS galaxy photometric sample is not
completed yet. Therefore, the luminosity function analysed in this
paper should be considered as a lower limit of the true cluster LF,
since we could miss low surface brightness galaxies especially at the
faint end. Bernstein et al. 1995, Ulmer et al. 1996 and Adami et
al. 2000 explore these issues in a series of papers on the faint LF of
the Coma cluster and conclude that LSB galaxies in Coma were
inconsequential. Moreover, Cross et al. 2004 compare the completeness
limits in magnitude and surface brightness of SDSS-EDR and SDSS-DR1
with the Millennium Galaxy Catalogue (MCG).  MCG is a deep survey with
limit in surface brightness 26 mag $arcsec^{-2}$.  They use the MCG
bright galaxies catalogue with galaxies in the magnitude range $16 \le
B \le 20$ (where $B=g+0.39(g-r)+0.21$ for DR1 magnitudes) for the
comparison with the SDSS-EDR-DR1 catalog.  They show that in the range
$21 \le \mu_e \le 25$ mag $arcsec^{-2}$ the incompleteness of SDSS-EDR
is less than 5\% and is around 10\% in the range $25 \le \mu_e \le 26$
mag $arcsec^{-2}$.  In the present work, for most of the clusters the
galaxies contributing to the DGR are faint galaxies in the magnitude
range $19 \le r \le 21$ mag . In this region of magnitude 65\% of the
objects lie at $\mu_e \le 23$ mag $arcsec^{-2}$, 30\% in the range $23
< \mu_e \le 24$ mag $arcsec^{-2}$, and 5\% at $mu_e \ge 25$ mag
$arcsec^{-2}$.  If we can apply also in this range of magnitude the
results of Cross et al. 2004, the incompleteness correction for low
surface brightness selection effects should be around 5\%. Therefore
we expect that the LSB galaxies do not contribute in a egregious way
to our luminosity function and cannot change significantly the DGR
calculated in the paper.

\begin{figure}
\begin{center}
\begin{minipage}{0.5\textwidth}
\resizebox{\hsize}{!}{\includegraphics{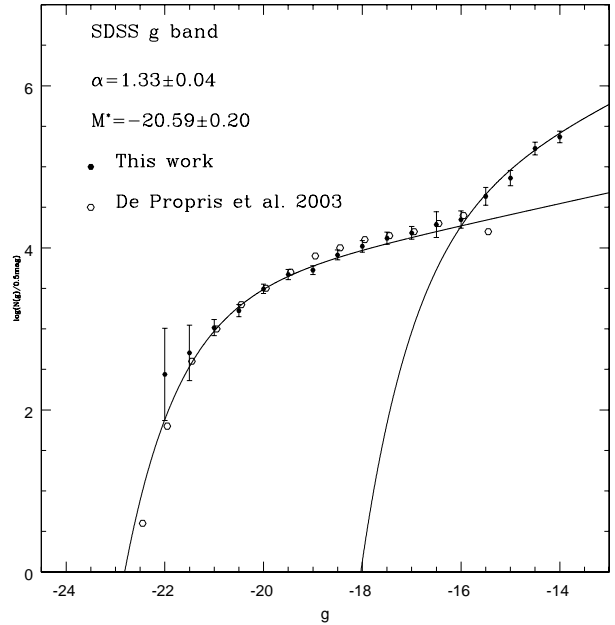}}
\end{minipage}
\end{center}
\caption{  
The plot shows the Composite LF calculated with the prescription of
Colless (1989) with a global background correction and within 1.5 Mpc
$\rm{h}^{-1}$ in the g band (filled points) and the De Propris et al.
(2003) Composite LF in the b band derived from the 2df spectroscopic
data (empty points).}
\label{deprop}
\end{figure}

\begin{table*}
\caption{ Schechter parameters fitted to the Composite LF retrieved in the literature.}
\begin{center}
\begin{tabular}[b]{cccccccccc}
\hline
Survey & &band & $m_{lim}$& $M^*$ & $\alpha$ & $\phi^*$ & evolution & Reference\\
 & & & &&&($\times 10^{-2}h^3$ $\rm{Mpc}^{-1}$) & correction& &  \\
\hline
AUTOFI           &           & $b_J$      & -14      &-19.30$\pm$0.13  & -1.16$\pm$0.05 &2.45$\pm$0.35& no &(1) \\
Stromlo-APM      &           & $b_J$      & -15      & -19.50$\pm$0.13 & -0.97$\pm$0.15 &1.40$\pm$0.17& no & (2) \\
SSRS2            &           & $m_{B(0)}$ & -14      & -19.45$\pm$0.08 & -1.16$\pm$0.07 &1.09$\pm$0.30& no & (3)\\
CfA2             &           & $m_Z$      & -16.5    & -18.8$\pm$0.3   & -1.0$\pm$0.2   &4.  $\pm$1   & no & (4) \\
EPS              &all        & $b_j$      & -12.4    & -19.61$\pm$0.06 & -1.22$\pm$0.06 &2.0 $\pm$0.4 & no &  (5)\\
                 &early type & $b_j$      & -12.4    & -19.62$\pm$0.09 & -0.98$\pm$0.09 &1.1 $\pm$0.2 & no &  \\
                 &late type  & $b_j$      & -12.4    & -19.47$\pm$0.10 & -1.40$\pm$0.09 &1.0 $\pm$0.2 & no &  \\
2dF              &all        & b          & -13      & -19.79$\pm$0.04 & -1.19$\pm$0.01 &1.59$\pm$0.14& no & (6) \\
                 &early  type& b          & -17      & -19.58$\pm$0.05 & -0.54$\pm$0.02 &0.99$\pm$0.05& no\\
                 &late type  & b          & -13      & -19.15$\pm$0.05 & -1.50$\pm$0.03 &0.24$\pm$0.02& no\\
2dF              &           & b          & -16.5    & -19.66$\pm$0.07 & -1.21$\pm$0.03 &1.61$\pm$0.08& yes & (7) \\
LCRS             &           & r          & -17.5    & -20.29$\pm$0.02 & -0.70$\pm$0.05 &1.9 $\pm$0.1 & no & (8) \\
CNOC2            &early type & $R_C$      & -17      & -20.50$\pm$0.12 & -0.07$\pm$0.14 &1.85$\pm$0.37& yes &  (9) \\
                 &late type  & $R_C$      & -16      & -20.11$\pm$0.18 & -1.34$\pm$0.12 &0.56$\pm$0.30& yes \\
SDSS             &           & $u^{0.1}$  & -15.54   & -17.93$\pm$0.03 & -0.92$\pm$0.07 &3.05$\pm$0.33& yes &(10) \\
DR1              &           & $g^{0.1}$  & -16.10   & -19.39$\pm$0.02 & -0.89$\pm$0.03 &2.18$\pm$0.08& yes \\
                 &           & $r^{0.1}$  & -16.11   & -20.44$\pm$0.01 & -1.05$\pm$0.01 &1.49$\pm$0.04& yes \\
                 &           & $r$        & -16.11   & -20.54          & -1.15          &1.77         & no \\
                 &           & $i^{0.1}$  & -17.07   & -20.82$\pm$0.02 & -1.00$\pm$0.02 &1.47$\pm$0.04& yes \\
                 &           & $z^{0.1}$  & -17.34   & -21.18$\pm$0.02 & -1.08$\pm$0.02 &1.35$\pm$0.04& yes \\
\hline
\end{tabular}
\label{field}
\end{center}
\parbox{18cm}{
Note. References: (1)Loveday et  al. (1992), (2)  Ellis et al. (1996),
(3) Marzke \& Da Costa (1997), (4) Marzke et al.  (1994), (5) Zucca et
al. (1997), (6) Madgwick et al. (2002), (7) Norberg et al. (2002), (8)
Lin et al. (1996), (9) Lin et al. (1999),  (10) Blanton et al. (2003).}
\end{table*}

\subsection{Comparison with previous work}
The results obtained by previous works are already shown in Table 1.
All the works in the literature analyse only the relatively bright end
of the cluster LF.  In fig. \ref{deprop} is shown the very good
agreement between our results in the g band and the Composite LF of
De Propris et al.  (2003), which uses the 2df spectroscopic data to
define the cluster membership.  It is clear in the figure that even
the 2df composite LF, which is supposed to be very deep, does not cover
the dwarf galaxy region analysed in this work. Therefore, we can 
compare only our bright end composite LF with the results found in the
literature.

We consider mainly the luminosity function in the g band since there
is a large number of b and g band cluster composite LFs in the
literature to compare with. After correcting the absolute magnitudes
for the different cosmology and for colors, our $M_g^*$ perfectly
agrees with almost all the the previous results except for Goto et
al. 2002. The disagreement with this work is larger than 3
$\sigma$. The reason of the discrepancy with the previous work
obtained with SDSS data is ascribable to the different quality of the
photometry between the last data release (DR2), used in this work, and
the Early Data Release (EDR) used in Goto et al. (2002).

The slopes of the Composite LF in the g band retrieved in literature
lie in a very large range of values from -1.50 to -0.94. Therefore,
there seems to be not an overall agreement in the literature about the
slope of the cluster composite LF.  Nevertheless, several of the works
retrieved in the literature should not be taken into account in this
comparison. In fact, the results of Garilli et al.  (1999) and
Paolillo et al.  (2001) should be excluded from our analysis , since
we notice in a previous paragraph that their results depend on the
method applied to derive the composite LF. Moreover, we would exclude
from our analysis also the results of Goto et al. 2002 for two
different reasons. First, Goto et al. 2002 uses a different SDSS
dataset (EDR) with lower quality in the photometry. Secondly, in that
work the background is calculated locally in an annulus around the
cluster center with outer radius of 1.3 Mpc $\rm{h}^{-1}$ and inner
radius of 1.0 Mpc $\rm{h}^{-1}$.  Since the background is calculated
within the cluster region (within an Abell radius of 1.5 Mpc), where
the fraction of cluster galaxies could be very high, such background
correction would subtract to a substantial degree the contribution of
that cluster galaxy population from the individual and the Composite
LF.  This suspicious background subtraction could explain the very
flat LF obtained by Goto et al. 2002 in all the Sloan wavebands.  In
conclusion, if we exclude the works of Garilli et al.  (1999),
Paolillo et al.  (2001) and Goto et al.  (2002) from our analysis, the
range of $\alpha$ is reduced significantly to the values between -1.50
and -1.22. All the values of the slope of our g band composite LF
perfectly fit in this range of results.

\subsection{Comparison with the field}

One of the most important and interesting aspects of the luminosity
function is the comparison of the LFs derived in different
environments.  The SDSS field luminosity function is given by Blanton
et al. (2003).  In this work the absolute magnitude limit is around
$-16+5log(h)$ in the g and r bands and $-17+5log(h)$ in the i and z
bands. Therefore, the field LF is not studied in the magnitude range
of the dwarf galaxies.  We compare, then, only the bright end of the
cluster luminosity function with the SDSS field LF. The result of the
comparison with the field LF of Blanton et al. (2003)is that the field
LF is systematically flatter than the cluster LF in any band, while
the cluster $M^*$ is brighter than the field $M^*$ of about 0.5
mag. However, it is important to notice that there is not an overall
agreement in the literature about the values of slope and $M^*$ in the
field luminosity function. As Table \ref{field} shows, most of the
results reveal a very poor agreement only within 3 $\sigma$, while
several values (see, e. g., the CfA2 LF of Marzke et al.  1994) do not
agree at all with the results of the other surveys. For example, if we
compare our cluster lf with the 2df field lf, we should conclude, in
agrrement with De Propris et al. 2002, that the slope of the cluster
lf is quite consistent with the field lf, while the characteristic
magnitute is about 0.5 mag brighter than the field $M^*$. Therefore,
we have to conclude that the not good agreement found generally in the
literature does not allow us to a conclusive comparison between the
luminosity function of different environments.

Since the magnitude range of our faint end cluster LF is not covered
by the SDSS field LF, we have to compare our results with other
surveys.  Loveday 1997 in the Stromlo-APM survey finds that the number
of faint galaxies seen in projection on the sky is much larger than
expected for a flat faint-end Schechter function. Moreover, they show
that the best fit function for the field luminosity function is a
"double power-law" Schechter function. Lin et al.  (1996) finds in the
Las Campana Redshift Survey (LCRS) that the Schechter function is a
good approximation of the magnitude range $ -23 \le M_r -5log(h) \le
-15.5$ for the field LF, but there is a significant excess relative to
the Schechter fit at the faint end $M_r \ge -17.5$.  Zucca et al.
(1997) finds a steepening of the field LF at $M_{b_J} \le
-17.5+5log(h)$ from the ESO Slice Project (ESP) galaxy redshift
survey.  A Schechter function is an excellent representation of their
data points at $M_{b_J} \le -16 + 5log(h)$, but at fainter magnitude
it lies below all the points down to $M_{b_j} = -12.4 +5log(h)$.  They
find that the best fit to the data is a two-law fit given by a
Schechter function plus a power law with slope -1.5.  They conclude
that the faint end steepening is almost completely due to galaxies
with emission lines.  In fact, dividing the galaxies in two samples
(i.e.  galaxies with and without emission lines) they find very
significant differences in their luminosity functions.  Galaxies with
emission lines show a significantly steeper faint end slope and a
slightly fainter $M^*$.  However, it is noteworthy that in their
results the Schechter function is a inadequate fit especially for the
galaxies without emission lines, which show a significant evidence of
an upturn in the dwarf region, while the LF of galaxies with emission
lines is much more compatible with a steep Schechter function.  A
similar difference in the best fit parameters of galaxies with and
without emission lines has been found also in the LCRS, Lin et al.
1996, although for each subsample their best fit is significantly
flatter than the corresponding slope in the EPS survey.

A partially different result comes from the 2dF survey, which shows
for the first time significant evidence for the presence of a
substantial passive dwarf population.  In fact, Madgwick et al.
(2002) find that the Schechter function provides an inadequate fit of
the LF calculated over the magnitude range $-22 \le M_{b_J} -5log(h)
\le -13$, especially     for   the most passive   and     star-forming
galaxies. They conclude that a Schechter function is not a good fit to
the data over the entire $M_{b_J}$ magnitude range and that this is
mostly due to an overabundance of the faint passive star-forming
galaxies relative to the bright objects. In fact, their sample of
passive galaxies clearly show a very significant increase in the
predicted number density of faint galaxies.  Moreover, they argue that
the small size of the other surveys has meant that only a
statistically insignificant number of galaxies have contributed to the
faint magnitude range. Hence previous studies could not determine if
the features observed at the faint end were real or a consequence of
the small volume being sample at these magnitudes.

Our results are more in agreement with the results of the ESP survey
of Zucca et al. 1997. In fact, as mentioned in section 5.2, the faint
end of our clusters should be due to a significant number of very
faint late type galaxies in clusters, which should be compatible with
the emission line galaxies observed by Zucca et al.  1997.  There is
also a qualitative good agreement with the results of the 2dF survey
of Madgwick et al.  (2002), since the late type galaxies in their
sample seem to have a steeper LF than the early type galaxies, even if
they conclude that the incompatibility of a Schechter function with
the global field LF should be due to the passive galaxies.  However,
it is important to stress that we can compare only qualitatively our
cluster LF with the field LF retrieved in literature since we are
using different wavebands and we are covering different magnitude
range.  Therefore, a quantitative comparison between the different
environments requires absolutely the measure of the field luminosity
function in the Sloan waveband and in the faint magnitude region.

\section{Conclusion}
The  main conclusion of  our analysis  are as follows: 
\begin{itemize}
\item we determine the   composite LF of galaxies in   clusters
from the SDSS data.  The LF clearly shows a bimodal behavior with an
upturn and a evident steepening in the faint magnitude range in any
SDSS band.  The LF is well fitted by the sum of two Schechter
functions.  The results are well confirmed by different methods of
background subtraction. The observed upturn of the faint galaxies has
a location ranging from -16 +5log(h) in the g band to -18.5 +5log(h)
in the z band.
\item   The  bright end LF shows   the classical slope  of -1.25 in
each photometric band, while $M^*$ is brighter in the red bands than
in the blue bands.  The distribution of the Schechter parameters
obtained fitting only the bright end of the individual cluster LF is
close to a Gaussian around the corresponding value of the composite
bright-end LF.  We check the dependence of the Schecter parameters
of the composite LF on the clustercentric distance calculating the LF
within different cluster apertures.  We do not find any significant
variation of the results with different apertures.  Therefore, we
conclude that the bright-end of the galaxy clusters is universal in
different cluster environments, both in different systems and in
different locations within the same cluster.
\item The faint end  LF is much steeper than the bright  end  LF with 
slope $-2.5 \le \alpha \le -1.6$.  We apply different tests to check
whether the observed faint end in the single clusters is due to the
presence of background large scale structures or a second cluster on
the line of sight.  To check the first possibility we measure the
individual cluster LF with a color cut method to identify the cluster
members. We obtain the same slope observed with the statistical
background subtraction.  Moreover, we observe that the faint
population is dominated by galaxies with colors compatible with late
type galaxies at the cluster redshift.  We, then, conclude that the
observed steepening of the cluster LF is due to the presence of a real
population of faint cluster galaxies.

\item We defined the Dwarf  to Giant galaxy Ratio DGR as the ratio 
between the number of galaxies in the magnitude range $ -18 \le M \le
-16.5$ and the number of galaxies brither than -20 mag.  In each
waveband the DGR seems to slightly increase from the very center 0.3
Mpc $\rm{h}^{-1}$ to 1.0 Mpc $\rm{h}^{-1}$.  The distribution of DGR
of the single clusters has a peak around the value predicted by the
composite bright-end and and a large spread at larger values. We,
then, conclude that the faint end of the cluster LF is not universal
and that the fraction of dwarf galaxies varies from cluster to
cluster. We check the relation between the DGR and the cluster
richness and between DGR and $M^*$ through the Spearman's rank
coefficient and we do not find any correlation between the parameters.
\end{itemize} 
We compare the above results with the field LF calculated in the
Sloan Digital Sky Survey and in other surveys.  The magnitude range
covered by the SDSS field LF allows us to compare only the bright end
of the cluster luminosity function with the field LF.  Unfortunately
there is no good agreement between the results retrieved in the
literature.  Therefore we cannot perform a conclusive comparison
between the LF of the different environments.  Moreover, several
surveys find evidence for the presence of an upturn at the faint end
of the field luminosity function in agreement with our results for the
cluster LF.  In particular Zucca et al. (1997) find in the ESP field
LF evidence for the presence of a late type galaxy population
dominating the faint end of the field luminosity function.  However,
it is important to stress that this is only a qualitative comparison
and does not allow us to any conclusion about the nature of the faint
population in clusters and in the field.  We can only conclude that
the faint end of the cluster LF is systematically steeper than the
field LF, although the field LF seems to show some evidences for an
excess of galaxies in the faint magnitude range relative to a
Schechter function.

Hierarchical clustering theories of galaxy formation generically
predict a steep mass function of galactic halos (Kauffmann, White \&
Guideroni 1993; Cole et al. 1994). This is in conflict with the flat
galaxy LF measured in the field and in diffuse local groups, but not
with the steep LF measured in many clusters.  However in the
hierarchical universe, clusters form relatively recently from the
accretion of smaller systems. The dynamical processes that operate in
clusters are destructive. Ram pressure stripping (e.g. Moore \& Bauer
1999) and gravitational tides/galaxy harassment (e.g.  Moore et
al. 1996, 1998) will both tend to fade galaxies by removing gas or
stripping stars. These processes are most effective for less massive,
less bound systems. Hence, we might expect to see a flattening of the
faint end slope in clusters compared to the field, rather than the
observed steepening.

Understanding the nature of the observed faint galaxy population
requires a more detailed study of the galaxy population in cluster
through the analysis of the morphological type, the colors and the
study of the relation between the fraction of dwarf galaxies and the
cluster parameters such as the cluster mass, velocity dispersion or the
X-ray luminosity.  Moreover, the origin of this faint population can
be understood only if a conclusive comparison between cluster and
field is possible. At the moment, as we discussed above, the SDSS
field LF based on the Sloan spectroscopic galaxy sample does not allow
to an exhaustive comparison and analysis of the different
environments.

It is clear from our results and from all the existing works in
literature that the composite bright end of the cluster LF can give
useful information on the global cluster properties (such as the total
optical luminosity, which is dominated by the very bright cluster
galaxies), but it does not provide useful information on the cluster
galaxy population as a whole. Equally it is clear that a Schechter
function is a good fit of the cluster LF only in a very restricted
magnitude range (the bright end). The photometric data available now
should make it possible to consider non-parametric comparisons between
the individual and the composite cluster LFs using the full range of
the available data. Our results on the dwarf galaxy fraction are the
first step in this direction, but it must be possible to devise a test
that does not require a split in bright and faint galaxies but
consider the cluster galaxy population as a whole.

\vspace{2cm}

Funding for the creation and distribution of the SDSS Archive has been
provided by the Alfred P.  Sloan Foundation, the Participating
Institutions, the National Aeronautics and Space Administration, the
National Science Foundation, the U.S.  Department of Energy, the
Japanese Monbukagakusho, and the Max Planck Society. The SDSS Web site
is http://www.sdss.org/. The SDSS is managed by the Astrophysical
Research Consortium (ARC) for the Participating Institutions.  The
Participating Institutions are The University of Chicago, Fermilab,
the Institute for Advanced Study, the Japan Participation Group, The
Johns Hopkins University, Los Alamos National Laboratory, the
Max-Planck-Institute for Astronomy (MPIA), the Max-Planck-Institute
for Astrophysics (MPA), New Mexico State University, University of
Pittsburgh, Princeton University, the United States Naval Observatory,
and the University of Washington.


\begin{thebibliography}{}
\bibitem[Abazajian et al.(2003)]{dr1}
Abazajian, K., Adelman, J., Agueros, M.,et al. 2003, AJ, 126, 2081 (Data Release One)
\bibitem[Adami et al.(2000)]{adami}
Adami, C., Ulmer, M. P., Durret, F. et al. 2000, A\&A, 353, 930
\bibitem[Beijersbergen et al.(2002)]{bei}
Beijersbergen, M.,Hoekstra, H., Van Dokkum, P.G. 2002, MNRAS, 329, 385
\bibitem[Bernstein et al.(1995)]{bernstein}
Bernstein, G. M., Nichol, R. C., Tyson, J. A. et al. 1995, AJ, 110, 1507
\bibitem[Blanton et al.(2001)]{blanton1}
Blanton, M. R., Dalcanton, J., Eisenstein, D., et al. 2001, AJ, 121, 2358
\bibitem[Blanton et al.(2003)]{blanton}
Blanton, M.R., Lupton, R.H., Maley, F.M. et al. 2003, AJ, 125, 2276 (Tiling Algorithm)
\bibitem[Blanton et al.(2003)]{blanton2}
Blanton, M.R., Hogg, D.W., Bahcall, N.A. et al., 2003, ApJ, 592, 819
\bibitem[B\"ohringer et al.(2000)]{bh1}
B\"ohringer, H., Voges, W.; Huchra, J. P., et al. 2000, ApJS, 129, 435
\bibitem[B\"ohringer et al.(2001)]{bh2}
B\"ohringer, H.,  Schuecker, P., Guzzo, L., et al. 2001, A\&A, 369, 826
\bibitem[B\"ohringer et al.(2002)]{bh3}
B\"ohringer, H., Collins, C. A., Guzzo, L., et al. 2002, ApJ, 566, 93
\bibitem[Boyce et al.(2001)]{boy}
Boyce, P.J., Phillips, S., Bryn Jones, et al., J. 2001, MNRAS, 328, 277
\bibitem[Cole et al.(1998)]{cole}
Cole, S., Aragon-Salamanca A., Frenk, C.S., et al. 1994,MNRAS,271,781
\bibitem[Colless et al.(1989)]{col}
Colless M. MNRAS, 237, 799
\bibitem[Cortese et al.(2003)]{cor}
Cortese L., 2003, A\&A,410L,25
\bibitem[Cross et al.(2004)]{cross}
Cross, N. J. G., Driver, S. P., Liske, J. et al. 2004, MNRAS, 349, 576
\bibitem[De Grandi et al.(1999)]{deg}
De Grandi, S., Böhringer, H., Guzzo, L., 1999, ApJ,514,148
\bibitem[De Propris et al.(2003)]{depropris}
De Propris, R., Colless, M., Driver, S. P., et al. 2003, MNRAS, 342, 725
\bibitem[Dressler et al.(1978)]{dre}
Dressler A., 1978,ApJ,223,765
\bibitem[Driver et al.(1994)]{dri}
Driver, S.P., Phillips, S., Davies, J.I. et al., 1994, MNRAS, 268,393
\bibitem[Ellis et al.(1996)]{ell}
Ellis, R.S., Colless, M., Broadhurst, T. et al.,  1996, MNRAS, 280, 235
\bibitem[Fukugita et al.(1995)]{fuk1}
Fukugita, M., Shimasaku, K.; Ichikawa, T. 1995, PASJ, 107,945
\bibitem[Fukugita et al.(1996)]{fuk2}
Fukugita, M., Ichikawa, T., Gunn, J. E. 1996, AJ, 111, 1748
\bibitem[Garilli et al.(2001)]{gar}
Garilli, B., Maccagni, D., Stefano, A. et al., 2001, A\&A, 342, 408
\bibitem[Goto et al.(2002)]{goto1}
Goto, T., Sekiguchi, M., Nichol, R. C., et al. 2002, AJ, 123, 1807
\bibitem[Goto et al.(2002)]{goto2}
Goto, T., Okamura, S., McKay, T. A.,  et al. 2002, PASP, 123, 1807
\bibitem[Gunn et al.(1998)]{gunn}
Gunn, J.E., Carr, M.A., Rockosi, C.M., et al 1998, AJ, 116, 3040 (SDSS Camera)
\bibitem[Hogg et al.(2001)]{Hogg}
Hogg, D.W., Finkbeiner, D. P., Schlegel, D. J., Gunn, J. E. 2001, AJ, 122, 2129
\bibitem[Horner(2001)]{horner}
Horner, D. 2001, PhD Thesis, University of Maryland
\bibitem[Kauffmann et al.(1993)]{kau}
Kauffmann, G., White, S.D.M., Guideroni, B. 1993, MNRAS, 264, 201
\bibitem[Kochanek et al.(2001)]{koch}
Kochanek, C. S., Pahre, M. A., Falco, E. E., et al. 2001, ApJ, 560,566
\bibitem[Lin et al.(1996)]{lin1}
Lin, H., Kirshner, R.P., Shectman, S.A. et al., 1996, ApJ, 464, 60
\bibitem[Lin et al.(1999)]{lin2}
Lin, H., Yee, H.K.C., Carlberg, R.G. et al., 1999, ApJ, 518, 533
\bibitem[Loveday et al.(1992)]{lov1}
Loveday, J., Peterson,B.A., Efstathiou, G., et al., 1992, ApJ, 390,338
\bibitem[Loveday (1997)]{lov2}
Loveday, J. 1997, ApJ, 489,29
\bibitem[Lugger et al.(1986)]{lug1}
Lugger, P.M. 1986,ApJ,303,535
\bibitem[Lugger et al.(1989)]{lug}
Lugger, P.M. 1989, ApJ, 343, 572
\bibitem[Lumsden et al.(1997)]{lumsden}
Lumsden, S. L.,  Collins, C. A., Nichol, R. C.,et al. 1997, MNRAS, 290, 119
\bibitem[Lupton et al.(1999)]{lup1}
Lupton, R. H., Gunn, J. E., Szalay, A. S. 1999, AJ, 118, 1406
\bibitem[Lupton et al.(2001)]{lup2}
Lupton, R., Gunn, J. E., Ivezi\'c, Z.,  et al.  2001, in ASP Conf. Ser. 238, Astronomical Data Analysis Software and Systems X, ed. F. R. Harnden, Jr., F. A. Primini, and H. E. Payne (San Francisco: Astr. Soc. Pac.), p. 269 (astro-ph/0101420)
\bibitem[Madgwick et al.(2002)]{mad}
Madgwick, D.S., Lahav, O., Baldry, I.K. et al., 2002, MNRAS, 333, 133
\bibitem[Marzke et al.(1994)]{mar}
Marzke, R.O., Huchra, J.P., Geller, M.J., 1994, ApJ, 428,43
\bibitem[Marzke et al.(1994)]{mar1}
Marzke, R.O. \& Da Costa, L.N., 1997, AJ, 113, 185
\bibitem[Moore et al.(1996)]{moo}
Moore, B., Katz, N., Lake, G., et al. 1996, Nat,379,613
\bibitem[Moore et al.(1998)]{moo1}
Moore, B., lake, G., Katz, N. 1998, ApJ, 495,139
\bibitem[Mulchaey et al.(2003)]{mul}
Mulchaey, J.S., Davis, D. S., Mushotzky, R. F.; Burstein, D. 2003,ApJSS, 145, 39
\bibitem[Norberg et al.(2002)]{nor}
Norberg, P., Cole, S., Baugh, C.M. et al., 2002, MNRAS, 336, 907
\bibitem[Paolillo et al.(2001)]{pao}
Paolillo, M., Andreon, S., Longo, G. et al., A\&A, 367, 59
\bibitem[Phillips et al.(1998)]{phi}
Phillips, S., Driver, S.P., couch, W.J. et al., 1998, ApJ, 498, L119
\bibitem[Popesso et al.(2004)]{pop}
Popesso, P., B\"ohringer, H., Brinkmann J., et al. 2004, A\& A, 423, 449 
\bibitem[Retzlaff(2001)]{retzlaff}
Retzlaff, J. 2001,XXIst Moriond Astrophysics Meeting, March 10-17, 2001 Savoie, France. Edited by D.M. Neumann  J.T.T. Van.
\bibitem[Rauzy et al.(2001)]{rau}
Rauzy, S., Adami, C., Mazure, A. et al., 1998, A\&A, 337, 31
\bibitem[Sabatini et al.(2003)]{sab}
Sabatini, S., Davies, J., Scaramella, R. et al., 2003, MNRAS, 341, 981
\bibitem[Schechter (1976)]{sche}
Schechter, P., 1976, ApJ,203,297
\bibitem[Schlegel et al.(1998)]{schlegel}
Schlegel, D., Finkbeiner, D. P., Davis, M. 1998, ApJ, 500, 525
\bibitem[Shimasaku et al.(2001)]{shimasaku}
Shimasaku, K., Fukugita, M., Doi, M., et al. 2001, AJ, 122, 1238
\bibitem[Smith at al.(1997)]{smi}
Smith, R.M., Driver, S.P., Phillips, S. et al., 1997, MNRAS, 287, 415
\bibitem[Smith at al.(2002)]{smith}
Smith, J.A., Tucker, D.L., Kent, S.M., et al. 2002, AJ, 123, 2121
\bibitem[Stoughton et al.(2002)]{stoughton}
Stoughton, C., Lupton, R.H., Bernardi, M., et al. 2002, AJ, 123, 485
\bibitem[Strauss et al.(2002)]{strauss}
Strauss, M. A., M.A., Weinberg, D.H., Lupton, R.H. et al. 2002, AJ, 124, 1810
\bibitem[Trentham et al.(1998)]{tre}
Trentham, N. 1998, MNRAS, 295,360
\bibitem[Ulmer et al.(1996)]{Ulmer}
Ulmer, M. P., Bernstein, G. M., Martin, D. R. et al. 1996, AJ, 112, 2517
\bibitem[Valotto et al.(1997)]{val}
Valotto, C., Nicotra, M.A., Muriel, H. et al., 1997, ApJ, 479, 90
\bibitem[Valotto et al.(2001)]{val2}
Valotto, C., Moore, B., Lambas, D. , 2001, ApJ, 479, 90
\bibitem[Valotto et al.(2004)]{val3}
Valotto, C., Muriel, H., Moore, B., et al., 2004, ApJ, 479, 90
\bibitem[Yagi et al.(2002)]{yag}
Yagi, M., Kashikawa, N., Sekiguchi, M. et al., 2002 AJ, 123, 87
\bibitem[Yasuda et al.(2001)]{yasuda}
Yasuda, N., Fukugita, M. Narayanan, V. K. et al. 2001, AJ, 122, 1104
\bibitem[York et al.(2000)]{york}
York, D. G., Adelman, J., Anderson, J.E.,  et al. 2000, AJ, 120, 1579
\bibitem[Zucca et al.(1997)]{zuc}
Zucca, E., Zamorani, G., Vettolani, G. et al., 1997, A\&A, 326, 477
\end{thebibliography}
\end{document}